\documentclass[12pt,a4paper]{article}
\usepackage{jheppub,mathrsfs}
\usepackage{amsmath,amssymb,amsfonts,mathtools,amsmath,amssymb,amsthm,amscd} 
\usepackage{float}
\usepackage{breqn}
\usepackage{array}
\usepackage{enumerate}
\usepackage{caption}
\usepackage{wrapfig}
\usepackage{fancybox}
\newcommand{\bea}{\begin{eqnarray}\displaystyle}
\newcommand{\eea}{\end{eqnarray}}

\newlength{\arrow}
{\setlength{\fboxsep}{15pt}
\setlength{\mylength}{\linewidth}%
\addtolength{\mylength}{-2\fboxsep}%
\addtolength{\mylength}{-2\fboxrule}%
\Sbox
\minipage{\mylength}%
\setlength{\abovedisplayskip}{0pt}%
\setlength{\belowdisplayskip}{0pt}%
\equation}%
{\endequation\endminipage\endSbox
\[\fbox{\TheSbox}\]}

\newdimen\tableauside\tableauside=1ex
\newdimen\tableaurule\tableaurule=0.8pt
\newdimen\tableaustep
\def\phantomhrule#1{\hbox{\vbox to0pt{\hrule height\tableaurule width#1\vss}}}
\def\phantomvrule#1{\vbox{\hbox to0pt{\vrule width\tableaurule height#1\hss}}}
\def\sqr{\vbox{%
  \phantomhrule\tableaustep
  \hbox{\phantomvrule\tableaustep\kern\tableaustep\phantomvrule\tableaustep}%
  \hbox{\vbox{\phantomhrule\tableauside}\kern-\tableaurule}}}
\def\squares#1{\hbox{\count0=#1\noindent\loop\sqr
  \advance\count0 by-1 \ifnum\count0>0\repeat}}
\def\tableau#1{\vcenter{\offinterlineskip
  \tableaustep=\tableauside\advance\tableaustep by-\tableaurule
  \kern\normallineskip\hbox
    {\kern\normallineskip\vbox
      {\gettableau#1 0 }%
     \kern\normallineskip\kern\tableaurule}%
  \kern\normallineskip\kern\tableaurule}}
\def\gettableau#1{\ifnum#1=0\let\next=\null\else
\squares{#1}\let\next=\gettableau\fi\next}

\begin{document}

\title{Higher Genera Perturbative/Non-Perturbative Relations}
\author{Emir O\u{g}uz Kaya, Eray Konya, Can Koz\c{c}az$^\dagger$}
\affiliation{Department of Physics, Bo\u{g}azi\c{c}i University, Istanbul, Turkey}
\affiliation{$^\dagger$Feza G\"{u}rsey Center for Physics \& Mathematics, Bo\u{g}azi\c{c}i University, Istanbul}
\emailAdd{emir.kaya2@std.bogazici.edu.tr,eray.konya@std.bogazici.edu.tr}
\emailAdd{can.kozcaz@bogazici.edu.tr}
\abstract{We study perturbative/non-perturbative (P/NP) relations for systems with several independent perturbative actions. The requirement that the quantum residues remain fixed leaves a choice of paths in parameter space. For each path, keeping all perturbative quantum periods fixed locally determines how the internal parameters must vary. The Riemann bilinear identity then gives a multi-parameter family of P/NP relations expressing the response of each dual non-perturbative period in terms of perturbative data. For a general sextic potential with three wells, the two outer-well actions determine how both tunneling periods vary. We also derive a flow equation for the quantum period matrix of the same potential. For a rational potential of the same genus, the residue at infinity fixes the ratio of the energy to the WKB coupling. Finally, we extend the construction to a third-order differential equation with a non-hyperelliptic curve of genus three.}

\maketitle
\section{Introduction}

Non-perturbative effects can determine basic properties of a quantum system, such as the lifetime of a metastable state or the splitting of otherwise degenerate energy levels. Their calculation is usually more involved than perturbation theory around a minimum. In a symmetric double well, for example, the perturbative expansion gives the same energy for the two states in each doublet to every order. One needs the action of a trajectory that crosses the barrier and the fluctuations about that trajectory to calculate their splitting semiclassically. An early indication that perturbation theory itself contains information about tunneling came from the work of Bender and Wu on the quartic anharmonic oscillator~\cite{Bender:1969si}. Upon continuation to negative quartic coupling the well becomes unstable, and the rate at which the particle escapes determines the leading large-order growth of the perturbative coefficients at positive coupling.

In the language of resurgence \cite{Ecalle:1981,DDP,Voros1983}, the large orders of the expansion about one saddle contain information about the first few fluctuations about other saddles~\cite{Aniceto:2013fka,Aniceto:2018bis}. In the double well the one-instanton contribution gives the leading splitting, while repeated tunneling gives further exponentially small corrections. Each contribution has its own fluctuation expansion. Together with perturbative expansions around saddles these form the sectors of the energy transseries~\cite{Zinn-Justin:2004vcw,Zinn-Justin:2004qzw,Jentschura:2010zza,Codesido:2017jwp,vanSpaendonck:2023znn}, and their resurgent structure is also reflected in the quantum periods~\cite{Gu:2022fss}.

We can recursively solve for the energy transseries using the quantization condition, which is the generalization of the Bohr-Sommerfeld quantization condition to take into account the effects beyond what the perturbation theory can capture. The  quantization condition is expressed in terms of two types of functions that either encodes the perturbative $B(E,g)$ or non-perturbative $A(E,g)$ data, including the quantum fluctuations, as a function of energy, coupling constant and the deformation parameters of the potential. The perturbative condition $B(E,g)=N+\tfrac12$ gives the common energy of the doublet as a series in the semiclassical parameter $g$. The instanton function $A(E,g)$ enters the tunneling factor and accounts for fluctuations about the classical instanton action. In~\cite{Alvarez:2000,Alvarez2004,Dunne:2013ada, Dunne:2014bca, Basar:2017hpr}, it was argued that for several one-dimensional potentials these functions satisfy a differential relation. In contrast to the usual resurgent relations between large and low orders of expansion terms around different saddles, this new type of relations allow one to determine low orders of the fluctuations about the instanton already in terms of the low orders of the inverse series $E(B,g)$. The relation between expansions at comparable orders has been known as the perturbative/non-perturbative (P/NP) relation.  

A useful way to understand the P/NP relation is to ask how the non-perturbative WKB period changes when the coupling varies and the perturbative action is held fixed~\cite{Gorsky:2014lia,Kozcaz:2026isf}. For the deformed anharmonic oscillator and the Jacobi elliptic potential, varying the shape of the potential brings in one additional piece of information~\cite{Cavusoglu:2023bai,Cavusoglu:2024usn}. Under continuation in a deformation parameter the contour defining the dual period can acquire a loop around a pole of the WKB differential. The resulting shift is proportional to the residue at that pole. Keeping the quantum residue fixed under such deformations makes the variation of the period independent of this ambiguity. The observation in \cite{Cavusoglu:2024usn} led to the Whitham construction of Ref.~\cite{Kozcaz:2026isf}, in which one chooses a residue-preserving flow and adjusts the energy to keep the perturbative quantum period fixed.

In \cite{Gahramanov:2015yxk,Gahramanov:2016xjj}, it has been argued that the simple form of the P/NP does not work for quintic or sextic potentials, which give rise to genus-two WKB curves, although it does for the cubic and quartic potentials, whose WKB curve is of genus-one. In this work, we study what the P/NP relation determines when there are several independent perturbative actions, hence generalizing the construction of~\cite{Gorsky:2014lia,Kozcaz:2026isf} to include potentials that give rise to WKB curve of genus $h$ bigger than one. In a three-well potential, for example, a change of energy that restores the left-well action after varying the coupling need not restore the right-well action. We allow appropriate coefficients of the potential to vary along with the energy. Locally, the perturbative periods determine these variations, provided their Jacobian with respect to the chosen parameters is invertible. The Riemann bilinear identity then determines how each dual period changes along the same flow. To recover a dual period from its variation one also needs its value on a reference slice. \cite{Kozcaz:2026isf} already reported that the approach to obtain the genus-one P/NP relation can straightforwardly be generalized to a higher-genus prescription and gave a P/NP relation for the generic sextic potential without derivation and checks. For related results on higher-genus polynomial oscillators we refer to \cite{Fischbach:2018yiu,Ture:2026ktb} and the two upcoming papers \cite{Gah,AT}. 

Keeping the quantum residue fixed leaves some freedom in how we deform the potential. For the general sextic, for example, we may vary the cubic and quartic coefficients independently as we change the coupling, while the residue condition fixes the variation of the quadratic coefficient. Choosing how the external parameters vary amounts to choosing a path in parameter space, with tangent $\mathcal{V}$. Once this choice is made, the requirement that all perturbative actions $B_I$ remain fixed determines the lift $\widehat{\mathcal{V}}$ uniquely wherever the period Jacobian is invertible. We thus obtain a multi-parameter family of P/NP relations. Each member of this family expresses the directional derivative $\widehat{\mathcal{V}}\mathcal{A}_I$ in terms of perturbative data and tells us how the same dual period responds when we deform the potential along a different direction.

We would like to present our method of constructing P/NP relations for different type of potentials, each of which we hope demonstrate a different aspect of possible potentials our approach could handle. Here is a list of our examples:

\begin{enumerate}[\bfseries i.]

\item{\textit{\textbf{The general sextic potential}}}

A sextic polynomial potential gives rise to a genus-two WKB curve and may have three wells, which allows us to ask whether perturbation theory in the two outer wells can determine the response of two inequivalent tunneling barriers. Reflection symmetry would identify the two barriers. We therefore take the general centered sextic $V(x)-\xi=x^6+\sum_{j=0}^4a_jx^j$, where $\xi$ is the energy. The constant and linear coefficients $a_0, a_1$ vary independently and can keep both outer-well actions fixed. For example, let $g$ vary at fixed $a_3,a_4$ and fixed scaled residue, and denote the resulting derivative with respect to $\log g$ at fixed actions by $\widehat{\mathcal V}$. The two P/NP equations for this flow are
\begin{equation}
 8g^2\,\widehat{\mathcal V}\mathcal A_I
   = a_4\,\frac{\partial a_0}{\partial B_I}
      +3a_3\,\frac{\partial a_1}{\partial B_I},
 \qquad I=L,R .
 \label{eq:intro-sextic}
\end{equation}
Both barrier periods are governed by the same functions of $B_L$ and $B_R$. The action derivative selects the barrier, with the other action and the external parameters held fixed. On the symmetric Chebyshev specialization the two equations reduce to the known genus-one relation on the elliptic quotient~\cite{Basar:2017hpr}. We also recover the functions $B_6$ and $A_6$ for the sextic anharmonic oscillator~\cite{Jentschura:2010zza}. With two actions, one can also ask how a change in one well affects the dual period associated with the other. This dependence is measured by an off-diagonal entry of the quantum period matrix. Its evolution along a residue-preserving Whitham time $T$ follows by differentiating the P/NP equation with respect to a second action. When the bilinear pairings of the internal variations vanish, the dual-period flows are action derivatives of a common function $\mathscr{H}_T$, and the result takes the form
\begin{align}
 \left.\partial_T\tau^{\rm }_{IJ}\right|_{\boldsymbol B}
   =\frac{i}{2\pi}\,
      \frac{\partial^2\mathscr{H} }{\partial B_I\,\partial B_J}.
\end{align}

This argument applies at higher genus as well. At leading WKB order the matrix is the ordinary Riemann period matrix, whose variation admits a branch-point residue formula~\cite{Rauch,Eynard:2007kz}. A quantum period matrix has also been studied in the sectorwise approach to the $\beta$-ensemble~\cite{Chekhov:2010zg}.

\item{\textit{\textbf{The quantum-corrected sextic}}}

We call potentials with terms including Planck's constant $\hbar$ quantum corrected potentials. They arise in quantum mechanical systems with Grassmann valued degrees of freedom after we ``integrate out'' those degrees of freedom, which also amounts to projecting to fixed fermion number Hilbert subspaces. In other words, the parameter controlling the fluctuations can also enter the potential itself. For the sake of presentation, we take the potential of the $U(z,g)=W(z)^2+\zeta g\,W'(z)+vz$, where $W(z)=z^3-az-b$, as in the class of quantum deformations studied in~\cite{Kozcaz:2016wvy,Kamata:2021jrs} (these are generalization of supersymmetric quantum mechanical systems to more than one Grassmann valued degree of freedom). The term proportional to $\zeta$ disappears from the classical potential, yet it shifts the middle-well action by $3\zeta/2$ relative to the sum of the outer-well actions. Its effect therefore survives in the quantum residue. Writing $u$ for the energy and $v$ for the tilt, we find
\begin{align}
 \left.g\partial_g\mathcal A_I\right|_{\boldsymbol B,a,b,\zeta}
  =\frac{1}{4g^2}\left(a\,\frac{\partial u}{\partial B_I}
                       -3b\,\frac{\partial v}{\partial B_I}\right),
 \qquad I=L,R .
\end{align}
The parameter $\zeta$ enters through the perturbative functions $u$ and $v$, even though it does not appear in the two coefficients on the right. We test this dependence by computing the two instanton functions independently of the well expansions. We find agreement through $g^2$ for generic shapes.

\item{\textit{\textbf{A rational potential of genus two}}}

This example should help us to examine how much of the P/NP relation is fixed by the genus of the curve. We choose our next potential to be a rational one of the form $V(z)=z^2+(\alpha+\beta z)/(z^2+\gamma^2)$ gives another genus-two curve with six branch points. Four come from zeros of the classical momentum squared and two from the simple poles of the potential at $z=\pm i\ell$. Clearing the denominator again gives a sextic equation, but the action differential retains the distinction between the two types of branch point. The classical action differential is regular above each finite pole, although the higher WKB differentials can be singular there~\cite{Koike}. The non-zero quantum residues at infinity fix $\xi/g$, such that varying the coupling already fixes the change in energy. In addition, the energy appears as a parameter in the residue, hence not one of the holomorphic coordinates as in the other examples we study. The numerator coefficients $\alpha, \beta$ must adjust to preserve the two actions. For this flow at fixed $\gamma$, the P/NP relation reads
\begin{align}
 \widehat{\mathcal V}\mathcal A_I
   =\frac{1}{2g^2}\frac{\partial \alpha}{\partial B_I},\qquad I=L,R ,
\end{align}
with $\widehat{\mathcal V}$ again normalized by $\widehat{\mathcal V}g=g$. We verify both components at orders $g^{-1},g,g^3$ with all parameters symbolic. Thus two potentials of the same genus assign different roles to the energy in the P/NP relation. The distinction is determined by the poles of the physical action differential.

\item{\textit{\textbf{The general octic potential}}}
The centered octic $V(x)-\xi=x^8+\sum_{j=0}^{6}a_jx^j$ gives three independent actions within ordinary quantum mechanics. One of its P/NP equations, for instance, is $2g^2\,\partial_{a_4}\mathcal A_I=-\partial_{B_I}a_2$, where $a_4$ varies at fixed actions and fixed remaining external data. Of the four external deformations preserving the residue, one combination is a dilation that leaves every scaled period invariant. The other three give independent flows of the dual periods.

\item{\textit{\textbf{A non-hyperelliptic curve of genus three}}}

Starting at genus $h=3$, there exist non-hyperelliptic curves, for which the minimum degree of a non-constant map to the Riemann sphere is greater than two. We would like to test whether our construction requires the two momentum sheets of a Schr\"odinger problem or it applies to non-hyperelliptic curves as well. We can address this question at the same genus as the octic by considering the third-order equation
\begin{align}
 \bigl[\hbar^3\partial_z^3+v\hbar\partial_z+z^4+az^2+bz+c\bigr]\psi=0 .
\end{align}
At $v=0$ this is a quantum Seiberg--Witten curve of the $(A_2,A_3)$ Argyres--Douglas theory. Its WKB periods are known from the correspondence between ordinary differential equations and integrable models~\cite{Ito:2021sjo}. At $v=0$, the WKB periods of this genus-three curve were studied using the ODE/IM correspondence~\cite{Ito:2021sjo}. In particular, the first quantum correction to a period was expressed in terms of derivatives of a classical period on the same cycle. Here we allow $v$ to vary and ask how each of the three dual periods responds to a change in $a$ when all three perturbative actions are held fixed. Here the classical momentum has three branches and its smooth curve is a non-hyperelliptic plane quartic of genus three, including at $v=0$. The parameters $b$ and $c$ give two holomorphic deformations. Allowing $v$ to vary supplies the third, so that all three actions can be kept fixed. In fact, the P/NP equation for a change of $a$ takes the particularly simple form
\begin{align}
 \left.\frac{\partial\mathcal A_I}{\partial a}\right|_{\boldsymbol B,\hbar}
   =-\frac{1}{\hbar^2}\frac{\partial v}{\partial B_I},
 \qquad I=1,2,3 ,
\end{align}
through the orders considered here. The derivative on the right can survive at $v=0$, so restricting the curve before differentiating would miss the flow on that very slice. For the scaled periods we verify the coupled relations symbolically through $\hbar^5$ at generic $v$, and through $\hbar^{11}$ at the quartic example of~\cite{Ito:2021sjo} on the $v=0$ slice. To the best of our knowledge, the three fixed-action P/NP equations for this deformed non-hyperelliptic curve have not appeared previously. In this example the choice of three independent actions gives the P/NP relation a meaning directly on the quantum Seiberg--Witten curve, beyond its original interpretation in terms of wells and tunneling barriers.

\end{enumerate}

The paper is organized as follows. In Section \ref{Construction}, we give a prescription for constructing P/NP relations at genus $h$. In Section \ref{sextic}, we apply this prescription to the general sextic potential and derive a family of P/NP relations. We explain how, given their values on a reference slice, the non-perturbative quantum periods can be recovered by integrating the flow equations using perturbative data. We also derive the variation of the period matrix $\tau$ along the same flow and give a Hamiltonian description of the flow. We conclude the section by showing how the genus-two P/NP relation for the symmetric sextic reduces to the genus-one relation on its elliptic quotient. In Section \ref{octic} , we derive the P/NP relations for the general octic potential. In Section \ref{qcsextic}, we study a quantum-corrected sextic potential motivated by supersymmetric quantum mechanics. In Section \ref{rationalpot} , we consider a rational potential whose WKB curve has genus two and derive the flow equation governing the variation of its non-perturbative quantum periods along a chosen deformation. Finally, in Section \ref{nonhyper}, we extend the construction beyond the Schr\"{o}dinger equation by applying it to a third-order differential equation that quantizes a non-hyperelliptic curve of genus three. In  Appendix \ref{cancellation}, we overview a derivation showing that the quantum corrections to the bilinear pairing residue cancel. In Appendix \ref{periods}, we collect expansions of the perturbative and non-perturbative quantum periods of the potentials we studied as examples in the main text.

\noindent\hrulefill

\textit{
As we were completing this work, we became aware of two forthcoming papers, one by Atakan \c{C}avu\c{s}o\u{g}lu and Kemal Tezgin \cite{AT} and the other by Ilmar Gahramanov \cite{Gah}. We thank the authors for agreeing to submit their papers to the arXiv on the same day as ours.}

\noindent\hrulefill


\section{Construction of the P/NP Relation}\label{Construction}

We review and extend the construction of the P/NP relation as a Whitham flow to potentials giving rise to higher genera WKB curves. For higher genera potentials, the approach is a straightforward generalization of the construction of the genus-one P/NP relation \cite{Kozcaz:2026isf,Gorsky:2014lia}, which consists of three main steps. Let us start by reviewing those three steps and adopting them to obtain genus-$h$ P/NP relations:
\begin{enumerate}[I.]
\item{\textit{Peripheral-cycle shift}}

    Let $\Sigma$ be a smooth compact surface of genus-$h$ with a set of poles $\mathcal{P}$, and we denote by $\lambda$ the WKB-differential of the curve $\Sigma$. The surface $\Sigma$ of genus-$h$ has $2h$ one-cycles, which we can group as $\mathfrak{a}=(\mathfrak{a}_1,...,\mathfrak{a}_h)$ and $\mathfrak{b}=(\mathfrak{b}_1,...,\mathfrak{b}_h)$, with the symplectic pairing $\mathfrak{a}_I \cdot \mathfrak{b}_J=-\delta_{IJ}$, $\mathfrak{a}_I \cdot \mathfrak{a}_J=0$ and $\mathfrak{b}_I \cdot \mathfrak{b}_J=0$. We integrate the quantum corrected (scaled) momentum along the cycles $\mathfrak{a}_I$ to evaluate perturbative quantum periods denoted by $B_I$, and the same quantum corrected momenta when integrated along the $\mathfrak{b}_J$-cycles give rise to the non-perturbative quantum periods, $\mathcal{A}_J$: 
    \begin{equation}
        B_I=\frac{1}{2\pi}\oint_{\mathfrak{a}_I}\lambda, \qquad \mathcal{A}_J=\frac{1}{i}\oint_{\mathfrak{b}_J}\lambda,\qquad \lambda=\frac{p}{g}\,dz,
    \end{equation}
    respectively. Let us denote the residue of the WKB one-form $\lambda$ WKB at a pole $p\in \mathcal{P}$ by $\mathcal{R}_p$,
    \begin{equation}
        \mathcal{R}_p=\operatorname{Res}_{p}\,\lambda
    \end{equation}
    Peripheral cycles can change the lift of a compact homology class to the punctured curve, and hence may deform the corresponding periods.
    \begin{equation}
        \mathfrak{b}_I\mapsto\mathfrak{b}_I+\sum_p n_p \delta_p, \qquad \mathcal{A}_I\mapsto \mathcal{A}_I+2\pi i \mathcal{R}_p
    \end{equation}
    Consequently, the relation between the perturbative and non-perturbative periods should remain invariant under the contribution resulting from monodromies. In other words, the P/NP relation should be insensitive to the peripheral cycle shifts for any pole. The holomorphic moduli parameters, which by definition do not change the singular behavior of the WKB one-form near singularities, do not contribute to the residue. Hence, for any  arbitrary holomorphic moduli (internal) parameter $u$ of the curve, $\partial_u \mathcal{R}=0$. We should not restrict ourselves only to the holomorphic moduli parameters to find residue-preserving flows. We write residue-preserving flows using the non-holomorphic moduli parameters $t$'s, which we also call external parameters. Then we can construct different flows which preserve the residue $\mathcal{R}$:
    \begin{equation}
        \mathcal{V}_j=\sum_i v^{i}_j(t)\frac{\partial}{\partial t_i},\qquad \mathcal{V}_j\mathcal{R}=0,\quad \textrm{for $p\in \mathcal{P}$}
    \end{equation}

 In other words, the choice of the residue preserving map is in general not unique when we consider potentials giving rise to higher genera WKB curves. We will show how different choices end up producing independent flows.     
\item{\textit{Fixed perturbative action}}
    
    Given the perturbative and non-perturbative quantum periods, the energy levels are found by solving the quantization conditions recursively. The perturbative quantum period is basically the quantum corrected action evaluated in the classically allowed region. The quantum corrections are can be obtained order by order solving the Riccati equation. The contours used to evaluate the perturbative periods $\mathfrak{a}_I$ enclose two turning points bounding the classically allowed region of the chosen wells. The perturbative energy for the $N^{\textrm{th}}$ level associated to the chosen perturbative well, can be computed simply by imposing the Bohr-Sommerfeld quantization condition,
    \begin{equation}
        B=N+\frac12
    \end{equation}
    As the deformation parameters of the curve vary the perturbative quantum periods varies accordingly, and so does the energy. We determine the changes in the energy as a function of the deformation parameters as we keep $N$ fixed. This is equivalent of saying that the perturbative quantum period must remain constant under the flow generated by changing the parameters of the potential. However, it is obvious that the perturbative periods are not generally invariant under the residue-preserving flows that we previously constructed. In general, for a potential with genus $h$ WKB curve, there are $h$ holomorphic perturbative quantum periods and $h$ holomorphic coordinates, on which the residues cannot depend by definition. We have enough number of holomorphic directions to lift the residue preserving vector field to one that also keeps the perturbative periods invariant under the flow generated by it. The potential of interest may not originally depend on all holomorphic coordinates associated with the curve. In such cases, we will add the necessary terms to the potential, derive the corresponding P/NP relation and then take carefully the zero limit; in particular, we pay attention possible differentiation with respect to them before taking the limit. The next step therefore consists of lifting the residue preserving vector field $\mathcal{V}$ to the perturbative quantum period preserving vector field $\widehat{\mathcal{V}}$. Note that this lift will not spoil the residue condition because the residue does not depend on the coordinates that we use for the lift. We can think of the lifted flow as a covariant derivative with respect to a perturbative connection on the fibration of internal moduli (holomorphic coordinates) over external parameters. Define the holomorphic-period Jacobian $\mathcal{J}_{I \alpha}=\frac{\partial B_I}{\partial u_{\alpha}}$, where $u_{\alpha}$ are the internal parameters $\alpha=1,\dots,h$. For the rest of our discussion, we will assume that the Jacobian is invertible,
     \begin{equation}
        \widehat{\mathcal{V}}\equiv\mathcal{V}\big |_{B_1,\mathellipsis,B_h}=\mathcal{V}-\sum_{\alpha=1}^{h}\Gamma^{\alpha}(\mathcal{V})\frac{\partial}{\partial u_{\alpha}},
    \end{equation}
    which satisfies the following relation by its definition
        \begin{equation}
        \widehat{\mathcal{V}}B_I=0, \qquad \textrm{for all $I=1,\mathellipsis,h$} 
    \end{equation}
    In fact, we can solve for $\Gamma^{\alpha}(\mathcal{V})$ and explicitly write the lifted vector field as,
    \begin{equation}
        \hat{\mathcal{V}}=\mathcal{V}-\sum_{I=1}^{h}(\mathcal{V}B_I)(\mathcal{J}^{-1})^{\alpha I}\frac{\partial}{\partial u_{\alpha}}.
    \end{equation}

\item{\textit{Riemann bilinear identity}}

The final step to obtain the flow equation for the non-perturbative periods relies on the Riemann bilinear identity and the fact that half of the periods on the WKB curve vanish by the action of the vector field $\widehat{\mathcal{V}}$ by construction. By a careful choice of the one-forms to integrate, half of the summands in the Riemann bilinear identity can be made to vanish.

We denote the holomorphic variations of the WKB one-form by $\omega_\alpha\equiv\partial_{u_{\alpha}}\lambda$ and also define $\Omega\equiv\widehat{\mathcal{V}}\lambda$. The bilinear identity for these forms become 
    \begin{equation}
        \sum_{I=1}^h\underbrace{ \left(\oint_{\mathfrak{a}_I}\Omega\right)\left(\oint_{\mathfrak{b}_I}\omega_{\alpha}\right)}_{=0}-\left(\oint_{\mathfrak{a}_I}\omega_\alpha\right)\left(\oint_{\mathfrak{b}_I}\Omega\right)=2\pi i \sum_{p\in \mathcal{P}}\operatorname{Res}_p \left (\omega_\alpha\, d^{-1}\Omega \right ),
    \end{equation}
    since the perturbative periods are $\widehat {\mathcal{V}}$-invariant. We can rewrite the P/NP relation for a potential with a genus-$h$ WKB curve in the closed form as a set of linear equations,
    \begin{equation}\label{short_eqn}
    \boxed{  \widehat {\mathcal{V}}\mathcal{A}=-\left(\mathcal{J}^T\right)^{-1}\mathcal{I}}
    \end{equation}
    We call $\mathcal{I}$ the Riemann bilinear pairing residue. In short, we claim that P/NP relations consists of $h$ linear equation describing the flow of the non-perturbative quantum periods $\mathcal{A}_I$ as a response to the change of the holomorphic variables with respect to the perturbative periods $B_J$. We claim that the Eq. \eqref{short_eqn} is the generic form for general genus-$h$ P/NP relations. 
\end{enumerate}


\section{Sextic potential}\label{sextic}
We apply our recipe described in the previous section to derive the P/NP relation for the generic sextic polynomial potential, which was earlier reported in \cite{Kozcaz:2026isf}. After a translation and setting the coefficient of the sextic monomial to one, the Schr\"{o}dinger equation\footnote{We use a slightly different scaling and get rid off the factor of $1/2$ in the kinetic term because we keep the coefficients of the potential arbitrary and prefer to avoid factors of $\sqrt{2}$ in some of the quantities we compute.} can be brought to the following generic form,
\begin{align}
g^2\psi''+Q(z)\psi=0, \qquad\mbox{with}\qquad Q(z)=-(z^6+a_4 z^4+a_3 z^3+a_2 z^2+a_1 z+a_0).
\end{align}
This potential gives rise to a genus-two WKB curve described by the following equation as an embedding in $\mathbb{C}^2$,
\begin{align}
p_{0}^2=Q(z),
\end{align}
where $p_0$ denotes the classical momentum, before including the correction dictated by the formal WKB expansion. The  WKB differential $\lambda=\frac{p}{g}\,dz$ has poles at positive and negative infinities with the residue 
\begin{align}
\mathcal{R}_{\pm}=\mbox{Res}_{\pm\infty}\left(\frac{p}{g}\, dz\right)=\mp\frac{i}{2g}\left( a_2-\frac{1}{4}a_{4}^{2}\right)\equiv\mp\frac{i}{2g}\mu.
\end{align}
The residue is exact and does not get any quantum corrections because the $(2n)^{th}$ correction to momentum $p_{2n}(z)=\mathcal{O}(z^{3-8n})$ for $n\geq1$. Having found the residue at the poles of the WKB one-form, we determine a vector field $\mathcal{V}$ that keeps $\mathcal{R}$ invariant; equivalently for our sextic potential we have $d(\mu/g)=0$. On any part of a path where $g$ varies, we can use $g$ as our coordinate and write it directly as
\begin{align}
g\frac{d\mu}{dg}=\mu.
\end{align}
If we want an additive flow parameter $t$ measuring the amount of rescaling such that two successive changes give additive increments, then we are left with the choice
\begin{align}
t=\log\frac{g}{g_{*}},
\end{align}
up to a choice of the starting point for the flow and the overall scale. Equivalently, we can write $g(s)=g_{*}e^{t}$ and we have
\begin{align}
\frac{d}{dt}=g\partial_g. 
\end{align}

In fact, $g\partial_g$ generates an infinitesimal rescaling $g\mapsto e^\varepsilon g$ andtThis operator reads off the WKB order of each term,
\begin{align}
g\partial_g\!\left(g^{2n-1}\rho_{2n}\,dz\right) =(2n-1)\,g^{2n-1}\rho_{2n}.
\end{align}

We can solve for $a_2$ along the flow preserving the residue $\mathcal{R}$, 
\begin{align}
a_2=\frac{1}{4}a_4^2+g\mu.
\end{align}
Thus, for each new value of $g$, we may choose $a_4$ and $a_3$ independently. Once we make these choices, $a_2$ is determined for fixed residue. In particular, we can specify a path through the initial sextic by giving two functions $a_4(g)$ and $a_3(g)$. We can also determine, how fast these coefficients are changing along the flow parametrized by $t$ above. We define the corresponding rates by
\begin{align}
c_4\equiv\frac{da_4}{dt}=g\frac{da_4}{dg},\qquad c_3\equiv\frac{da_3}{dt}=g\frac{da_3}{dg}.
\end{align}
We can find the rate of change of $a_2(t)$ to be 
\begin{align}
\frac{da_2}{dt}=\mu+\frac{1}{2}c_4 a_4.
\end{align}
The external vector field $\mathcal{V}$ preserving the residue, $\mathcal{V}\mathcal{R}_{\pm}=0$, can be written in general as
\begin{align}
\mathcal{V}=g\partial_g+c_4\partial_{a_4}+c_3\partial_{a_3}+\left( \mu+\frac{1}{2}c_4 a_{4}\right)\partial_{a_2}.
\end{align}
First, for simplicity, we choose $c_3=0$, i.e. $a_3$ is kept constant along the flow and we set $c_4\neq0$. In general, the vector field $\mathcal{V}$ does not keep the perturbative periods $B_I$'s invariant, thus we need to lift it using the holomorphic compensation directions, $a_0$ and $a_1$ in our case. $a_0$ is essentially the energy, which has been used also in the genus-one potentials as the holomorphic variable, and $a_1$ is the second holomorphic variable controlling essentially the tilting of the potential. As previously mentioned, they do not change the singularity structure of the WKB one-form.  For generic potentials giving rise to genus two WKB curves, we have two independent perturbative periods, $B_{I}$'s, and two holomorphic direction, let us consider the general lift of $\mathcal{V}$ by 
\begin{align}\nonumber
\widehat{\mathcal{V}} &=\mathcal{V}+\zeta^{0}\partial_{a_0}+\zeta^{1}\partial_{a_1},\\\label{liftedflow}
&=g\partial_g+c_4\partial_{a_4}+c_3\partial_{a_3}+\left( \mu+\frac{1}{2}c_4 a_{4}\right)\partial_{a_2}+\zeta^{0}\partial_{a_0}+\zeta^{1}\partial_{a_1}
\end{align}
such that $\widehat{\mathcal{V}} B_I=0$. Using the quantum holomorphic-period Jacobian defined previously by 
\begin{align}
\mathcal{J}_{I\alpha}\equiv\partial_{a_{\alpha}}B_I,
\end{align}
we can solve for $\zeta^i$ and get 
\begin{align}
\widehat{\mathcal{V}}=\mathcal{V}-(\mathcal{V}B_I)(\mathcal{J}^{-1})^{\alpha I}\partial_{a_\alpha}.
\end{align}
If we want to write more explicitly, we can call $h_I\equiv \partial_{a_0}B_I$, $k_I\equiv \partial_{a_1}B_I$, $q_I\equiv \mathcal{V}B_I$ and the determinant of the Jacobian $\mathcal{J}$ by $\Delta_{\mathcal{J}}$,
\begin{align}\nonumber
\widehat{\mathcal{V}}&=g\partial_g+c_4\partial_{a_4}+\left( \mu+\frac{1}{2}c_4 a_{4}\right)\partial_{a_2}+\frac{1}{\Delta_{\mathcal{J}}}\left(k_1 q_2-k_2 q_1 \right)\partial_{a_0}\\
&+\frac{1}{\Delta_{\mathcal{J}}}\left( h_2 q_1-h_1 q_2\right)\partial_{a_1}
\end{align}
Two natural fields associated with the two holomorphic variables are given by 
\begin{align}
\mathcal{E}_1=\frac{1}{\Delta_{\mathcal{J}}}\left ( k_2\partial_{a_0}-h_2\partial_{a_1}\right),\qquad\mbox{and}\qquad\mathcal{E}_2=\frac{1}{\Delta_{\mathcal{J}}}\left (-k_1\partial_{a_0}+h_1\partial_{a_1}\right),
\end{align}
which satisfy
\begin{align}
\mathcal{E}_IB_J=\delta_{IJ}.
\end{align}
In terms of these new vector fields, the horizontal lift takes a very compact form,
\begin{align}
\boxed{\widehat{\mathcal{V}}=\mathcal{V}-(\mathcal{V}B_1)\mathcal{E}_1-(\mathcal{V}B_2)\mathcal{E}_2}
\end{align}
By construction, we have $\widehat{\mathcal{V}}B_I=0$ and $\widehat{\mathcal{V}}\mathcal{R}_q=0$. 

Now, we wish to study how the dual periods, i.e. the quantum periods associated with the non-perturbative periods, change under the Whitham flow: $\mathcal{W}_I\equiv\widehat{\mathcal{V}}\mathcal{A}_I$. We will use the one-forms that we previously defined in our recipe, $\Omega\equiv \widehat{\mathcal{V}}\lambda$ and $\omega_\alpha\equiv \partial_{a_\alpha} \lambda$ for $\alpha=1,2$. From these definitions, the following identities follow:
\begin{align}
\oint_{\mathfrak{a}_I}\Omega&=2\pi\,\widehat{\mathcal{V}}B_I=0,\\
\oint_{\mathfrak{a}_I}\omega_\alpha&=\partial_{a_\alpha}\left( 2\pi B_I\right),\\
\oint_{\mathfrak{b}_I}\Omega&=i\mathcal{W}_I.
\end{align}
The Riemann bilinear identity for a genus-two curve gives us
\begin{align}
\sum_{I=1}^{2}\underbrace{\left(\oint_{\mathfrak{a}_I}\Omega \right)\left(\oint_{\mathfrak{b}_I}\omega_{\alpha} \right)}_{=0}-\left(\oint_{\mathfrak{a}_I} \omega_{\alpha}\right)\left(\oint_{\mathfrak{b}_I}\Omega \right)=2\pi i \sum_{P}\mbox{Res}_{P}\left( \Phi \omega_{\alpha}\right),
\end{align}
where we locally defined the primitive, $d\Phi=\Omega$. As indicated in the equation above, the first summand vanishes by construction. Using our previous definitions, we can rewrite this equation as 
\begin{align}\label{generalPNP}
\sum_{I=1}^{2}\mathcal{J}_{I\alpha}\mathcal{W}_I&=-\sum_{P}\mbox{Res}_{P}\left( \Phi \omega_{\alpha}\right),\\
\mathcal{J}^{T}\mathcal{W}&=-\mathcal{I}
\end{align}
which is in fact the genus-two P/NP relation, now it is a matrix equation rather than being a single equation. We can easily compute the residue vector 
\begin{align}
 \mathcal{I}\equiv\sum_{P}\mbox{Res}_{P}\left( \Phi \omega_{\alpha}\right)&= \frac{1}{8g^2}\begin{bmatrix}
           2c_4-a_{4} \\
           -3a_{3} 
         \end{bmatrix},
  \end{align}
which does not receive any quantum corrections, see the Appendix \ref{cancellation} for the details. For the sextic potential in consideration, we can explicitly write the flow of the non-perturbative period
\begin{align}\label{sexticPNP}
\boxed{\widehat{\mathcal{V}}\mathcal{A}_I= \frac{1}{8g^2}\left[ (a_4-2c_4)\frac{\partial a_0}{\partial B_I}+3a_3 \frac{\partial a_1}{\partial B_I}\right],\qquad I=1,2}
\end{align} 
As it is common for all known genus-one P/NP relations, the energy is expressed as a function of the perturbative quantum period; now perturbative periods since we have two. At the same time, we need to invert the second holomorphic variable $a_1$ as well and express it as a function of both perturbative periods:
\begin{align}
a_0&=a_0(B_1,B_2;\eta),\\
a_1&=a_1(B_1,B_2;\eta),
\end{align}
where $\eta$ is a collective parameter for the remaining deformation parameters of the potential and the quantum residue, which we keep fixed (c.f. Section \ref{hamiltonian}). These functions actually constitute the local quantum mirror map, and their existence is guaranteed as long as the determinant of the quantum Jacobian is non-vanishing, $\mbox{det}\,\mathcal{J}\neq0$.


\subsection{Family of P/NP Flows}

Allowing a non-zero cubic deformation rate $c_3=\dot{a}_3$ replaces $3a_3$ by $3a_3-4c_3$ in the $a_1$ component of the residue pairing, yielding the following P/NP relation:
\begin{align}\label{PNPc3c4}
\widehat{\mathcal{V}}\mathcal{A}_I= \frac{1}{8g^2}\left[ (a_4-2c_4)\frac{\partial a_0}{\partial B_I}+(3a_3-4c_3) \frac{\partial a_1}{\partial B_I}\right],\qquad I=1,2
\end{align}

Note that the lifted vector field $\widehat{\mathcal{V}}$, Eq. (\ref{liftedflow}), depends on $c_3$ and $c_4$, which do not depend on the potential but our choice of parametrization of the flow. Hence, we have the freedom to choose them independently from each other implying that we have in fact a two-parameter family of P/NP flows. Equivalently, we can write the following three flow equations: 
\begin{align}
\widehat{\mathcal{V}}_g\mathcal{A}_I\equiv\Big(g\partial_g +\mu\partial_{a_2}+\zeta_{g}^{0}\partial_{a_0}+\zeta_{g}^{1}\partial_{a_1} \Big)\mathcal{A}_I&=\frac{1}{8g^2}\left[a_4\frac{\partial a_0}{\partial B_I}+3a_3\frac{\partial a_1}{\partial B_I} \right],\\
\widehat{\mathcal{V}}_\chi\mathcal{A}_I\equiv\Big( \partial_{a_4}+\frac{1}{2}a_4\partial_{a_2}+\zeta_{\chi}^{0}\partial_{a_0}+\zeta_{\chi}^{1}\partial_{a_1}\Big)\mathcal{A}_I&=-\frac{1}{4g^2}\frac{\partial a_0}{\partial B_I},\\
\widehat{\mathcal{V}}_3\mathcal{A}_I\equiv\Big(\partial_{a_3}+\zeta_{3}^{0}\partial_{a_0}+\zeta_{3}^{1}\partial_{a_1}\Big)\mathcal{A}_I&=-\frac{1}{2g^2}\frac{\partial a_1}{\partial B_I},
\end{align}
where we included the proper lifts in each case, since the lifts vary for different seeds. In fact, we can simply write for the left hand sides,
\begin{align}
\widehat{\mathcal{V}}_g\mathcal{A}_I=\left.g\frac{\partial \mathcal{A}_I}{\partial g}\right|_{B_{1,2}, \mathcal{R},a_4,a_3},\qquad\widehat{\mathcal{V}}_\chi\mathcal{A}_I=\left.\frac{\partial \mathcal{A}_I}{\partial a_4}\right|_{B_{1,2}, \mathcal{R},g,a_3},\qquad\widehat{\mathcal{V}}_3\mathcal{A}_I=\left.g\frac{\partial \mathcal{A}_I}{\partial g}\right|_{B_{1,2}, \mathcal{R},g,a_4}.
\end{align}

However, these flows are not independent of each other. The first clue comes from the fact that the right hand side of the P/NP relation in Eq. (\ref{PNPc3c4}) vanishes at 
\begin{align}\label{vanishingconditions}
a_4-2c_4=0,\qquad\mbox{and}\qquad 3a_3-4c_3=0,
\end{align}
which lead to a very special type of flow. Consider dilation of the coordinate by $z\mapsto sz$ and the coefficients by $a_j\mapsto s^{6-j}a_j$, then the Schr\"{o}dinger equation will keep its form if the coupling constant after the scaling is given by $g_s=s^4 g$, in other words our flow parameter satisfies $t=4\log s$ and gives
\begin{align}
c_j=\frac{da_j}{dt}=\frac{6-j}{4}a_j,
\end{align} 
which are precisely the conditions in Eq. (\ref{vanishingconditions}). In other words, the right hand side of the P/NP relation vanishes precisely when the flow corresponds to dilation generated by the operator,
\begin{align}
\widehat{\mathcal{V}}_{\textrm{dil}}=g\partial_g+\frac{a_4}{2}\partial_{a_4}+\frac{3a_3}{4}\partial_{a_3}+a_2\partial_{a_2}+\frac{5a_1}{4}\partial_{a_1}+\frac{3a_0}{2}\partial_{a_0}.
\end{align} 
The first four terms of $\widehat{\mathcal V}_{\textrm{dil}}$ give the external residue-preserving field on $(g,a_4,a_3,a_2)$. The last two give its fixed-action lift along $a_1,a_0$. Indeed, the full WKB momentum scales as $p(z)\mapsto s^3p(z)$ such that the WKB one-form $\lambda=p(z)\,dz/g$ remains invariant. Thus every quantum period along the flow is invariant, which also explains directly why the right hand side of the P/NP equation vanishes. Moreover, the dilation operator can be written explicitly as a linear combination of the vector fields above\footnote{It is obvious that the terms including the external parameters match on both sides. The equivalence of the holomorphic parts follows from the fact that the period Jacobian $\mathcal{J}$ is invertible and the fact that the action of these operators vanish for any period.},

\begin{align}
\widehat{\mathcal{V}}_{\textrm{dil}}=\widehat{\mathcal{V}}_g+\frac{a_4}{2}\widehat{\mathcal{V}}_\chi+\frac{3a_3}{4}\widehat{\mathcal{V}}_3.
\end{align} 

The action of this operator on \textit{any} period vanishes, hence, we can conclude that only two of the three sextic P/NP flow equations carry independent information. In fact, from the explicit form on these operators above, we see that 
\begin{align}
\widehat{\mathcal{V}}_g=-\frac{a_4}{2}\widehat{\mathcal{V}}_\chi-\frac{3a_3}{4}\widehat{\mathcal{V}}_3.
\end{align} 

\par{We want to reach the same conclusion from a slightly more general angle. Our construction starts with a choice of a vector field $\mathcal{V}$ that keeps the quantum residue of the WKB differential fixed. The sextic potential with more deformation parameters makes it apparent that the choice of the vector field is far from being unique; we can focus on other flows while staying on the leaf $\mathcal{R}=\mbox{(constant)}$. We can horizontally lift these other vector fields to generate new Whitham flows in the extended coefficient space $\mathcal{M}$ as we previously did. In this section, we would like to argue that there is in fact one more independent P/NP-like flow for the deformed sextic potential that we can obtain by starting with a different vector field. }

Suppose the external deformation space has local coordinates $T^A$, with $A=1,\mathellipsis,m$. For every such time, we have an external field 
\begin{align}
\mathcal{V}_A=\frac{\partial}{\partial T^A}. 
\end{align}
Again, these vector fields by themselves do not keep the perturbative periods constant, therefore we need the horizontal lift defined by 
\begin{align}
\widehat{\mathcal{V}}_A&=\left.\frac{\partial}{\partial T^A}\right|_{\mathbf{B}},\\
&=\mathcal{V}_A+\zeta_{A}^{0}\partial_{a_0}+\zeta_{A}^{1}\partial_{a_1},
\end{align}
where we emphasized with the superscript that the coefficients of the lift depends on the particular direction we choose on the fixed residue leaf. The lift ensures that $\widehat{\mathcal{V}}_A B_I=0$, and once more we can make use of the Riemann bilinear identity to get the flow for the dual (non-perturbative) period. Let us define $\mathcal{W}^{(A)}_I=\widehat{\mathcal{V}}_A \mathcal{A}_I$, and following the previous steps will lead us to 
\begin{align}
\mathcal{J}^{T}\mathcal{W}^{(A)}=-\mathcal{I}^{(A)}.
\end{align}
Let us try to elaborate this observation on the sextic potential and derive a new P/NP flow equation: we will choose our times as $(T^1,T^2,T^3)=(g,\chi,a_3)$. In addition to the vector field that we have already used $\mathcal{V}_s$, there are two more vector fields that will keep the quantum residue $\mathcal{R}$ fixed,
\begin{align}
\mathcal{V}_3=\partial_{a_3}, \qquad\mbox{and}\qquad \mathcal{V}_{\chi} =\partial_{a_4}+\frac{a_4}{2}\partial_{a_2}.
\end{align}
We first obtain the flow equation starting with the vector field $\mathcal{V}_3$. The singular variation starts as
\begin{align}
d\Omega_3=-\frac{i}{2g}t^{-2}dt+\mathellipsis,
\end{align}
and the residue in the bilinear relation can be computed as before
\begin{align}
\mathcal{I} =\frac{1}{2g^2}\begin{bmatrix}
           0\\
           1
         \end{bmatrix},
\end{align}
and the Hamiltonian is given by 
\begin{align}
\mathscr{H}_3=-\frac{1}{2g^2}a_1.
\end{align}
The new flow equation is therefore 
\begin{align}
\widehat{\mathcal{V}}_3\mathcal{A}_i=\partial_{B_i}\left( -\frac{a_1}{2g^2}\right).
\end{align}
We can repeat the same steps again for $\mathcal{V}_{\chi}$ and get the associated flow equation:
\begin{align}
\widehat{\mathcal{V}}_{\chi} \mathcal{A}_i=\partial_{B_i}\left(-\frac{a_0}{4g^2}\right).
\end{align}
Hence, we obtained the vector fields and the associated P/NP relations that we have obtained before starting from different choices of the keeping the residue fixed.


\subsection{Integrating the P/NP relation}
\par{The novelty of the P/NP relation is in its constructive power to produce the non-perturbative information purely from the perturbative one. In its simplest form, we could simply integrate it, }
\begin{align}
A(E,g)=-\int\frac{dg}{g}\left(\frac{S}{g} \frac{\partial E}{\partial B}+B\right).
\end{align}
Such an integral to generate the non-perturbative data can be used for the generalized P/NP relation when deformation parameters are introduced in the potential; but since the general P/NP is a partial differential equation instead of an ordinary differential equation, it had to be supplemented by some boundary data. For example, straightforward integration missed some terms in the non-perturbative function $A$ for the Jacobi elliptic potential. 

Note that 
\begin{align}
\widehat{\mathcal{V}}g=g,\qquad\widehat{\mathcal{V}}a_4=c_4,\qquad\widehat{\mathcal{V}}\mathcal{R}_q=0,\qquad\widehat{\mathcal{V}}a_3=0,\qquad\widehat{\mathcal{V}}B_i=0.
\end{align}
We would like to introduce a dimensionless flow parameter
\begin{align}
s\equiv\log \frac{g}{g_*}, 
\end{align}
which satisfies $\widehat{\mathcal{V}}s=1$. Likewise, we can define a new parameter instead of $a_4$ by
\begin{align}
\chi\equiv a_4-s,
\end{align}
such that $\widehat{\mathcal{V}}\chi=0$. After this change of parameter, we have five parameters that remain constant under the Whitham flow: $(B_1,B_2,\mathcal{R}_q,a_3,\chi)$, whereas along the flow we have (for simplicity, let us choose $c_4=1$)
\begin{align}
g(s)=g_*e^s,\qquad\mbox{and}\qquad a_4(s)=\chi+s.
\end{align}
In these new coordinates, the vector field $\widehat{\mathcal{V}}$ has a particularly simple form,
\begin{align}
\widehat{\mathcal{V}}=\left. \frac{\partial}{\partial s}\right|_{B_1,B_2,\mathcal{R},a_3,\chi}.
\end{align}
The remaining moduli are obtained from the quantum mirror map:
\begin{align}
a_0&=a_0(B_1,B_2;g(s),\mathcal{R},a_3,a_4(s)),\\
a_1&=a_1(B_1,B_2;g(s),\mathcal{R},a_3,a_4(s)).
\end{align}
In this new adapted coordinate $s$ along the flow, the Hamiltonian equation governing the flow of the dual period now becomes an ordinary differential equation:
\begin{align}
\frac{d\mathcal{A}_I}{ds}=\left. \frac{\partial \mathscr{H} }{\partial B_I}\right|_{B_{J\neq I},\mathcal{R}_q,a_3,\chi},
\end{align}
which we can write more explicitly as 
\begin{align}
\frac{d\mathcal{A}_I}{ds}=\frac{1}{8g(s)^2}\left[\left( a_4(s)-2\right)\frac{\partial a_0}{\partial B_I}+3a_3\frac{\partial a_1}{\partial B_I} \right].
\end{align}
The derivatives on the right are evaluated at the current point of the flow with $(B_1,B_2,\mathcal{R},a_3,\chi)$ kept fixed. Integrating this equation from $s_0$, some initial starting point for the flow, to an arbitrary point $s$:
\begin{align}
\mathcal{A}_I(B;s)=\mathcal{A}_I(B;s_0)+\frac{\partial}{\partial B_I}\int_{s_0}^{s}\mathscr{H}(B;s')\,ds' ,
\end{align} 
where we made use of the fact that $B_i$'s are constant along the Whitham flow. Thus the change in each nonperturbative period is the $B_i$-gradient of one integrated generating function 
\begin{align}
\mathscr{G}(B;s,s_0)\equiv\int_{s_0}^{s}\mathscr{H}(B;s')\,ds' 
\end{align}
We get the integrated form of the genus-two P/NP relation:
\begin{align}
\boxed{\mathcal{A}_I(B;s)-\mathcal{A}_I(B;s_0)=\frac{\partial \mathscr{G}(B;s,s_0)}{\partial B_I}}
\end{align}
This also shows why boundary data remains necessary, the P/NP equation determines the variation of  $\mathcal A_I$, but not the initial values $\mathcal{A}_I(B;s_0)$. This is the genus-two analogue of the boundary-function ambiguity in the genus-one P/NP equations at fixed scaled residue observed earlier in \cite{Cavusoglu:2024usn}.


\subsection{Flow of the Period Matrix}

The period matrix $\tau$ encodes the complex structure of a Riemann surface and is defined only in terms of what we call here the classical periods. In quantum mechanics, we are interested in computing the full WKB periods including all quantum corrections and for potential giving rise to higher genera curves, the P/NP relations is interpreted as a set of flow equations of the non-perturbative quantum periods under a vector field, which keeps the residues of the WKB differential and perturbative quantum periods invariant. In this section, we want to show that our P/NP equation can be recast as the response of the non-perturbative (dual) quantum periods to changes in the perturbative quantum periods.

We start with the lighting review of the ordinary period matrix. Suppose $(\mathfrak{a}_I,\mathfrak{b}_I)$ with $I=1,\mathellipsis,h$ is a symplectic basis and normalize the holomorphic differentials $\omega_J$ by 
\begin{align}
\oint_{\mathfrak{a}_I}\omega_J=\delta_{IJ}.
\end{align}
The ordinary period matrix is then defined by 
\begin{align}
\tau^{(0)}_{IJ}=\oint_{\mathfrak{b}_I}\omega_J,
\end{align}
where we used the superscript $\tau^{(0)}$ to denote that this matrix is what we consider as the zeroth order classical period in the WKB expansion. We defined our quantum period using the scaled WKB differential with some conventional pre-factors as,
\begin{align}
B_I=\frac{1}{2\pi}\oint_{\mathfrak{a}_I}\lambda, \qquad \mathcal{A}_I=\frac{1}{i}\oint_{\mathfrak{b}_I}\lambda.
\end{align}

Each genus $h$ curve has $h$ internal deformations, which we denote here by $u_1,\mathellipsis,u_h$ (for our sextic potential, they were $a_0$ and $a_1$, the coefficients, whose variations left the pole structure of the WKB one-form invariant), and we assume that the quantum holomorphic-period Jacobian $\mathcal{J}_{I\alpha}=\partial_{u_\alpha}B_I$ is invertible. We can write the normalized quantum variation forms by
\begin{align}
\Omega_J=\frac{1}{2\pi}\frac{\partial\lambda}{\partial B_J},\qquad \oint_{\mathfrak{a}_I}\Omega_J=\delta_{IJ},
\end{align}
where we keep the external variables, which change the singular behavior of the WKB one-form, fixed as we differentiate with respect to the perturbative quantum periods. Differentiation with respect to the perturbative periods is in fact aligned with our discussion on the P/NP relation as we also used it in our example of sextic potential. In other words, we perform a coordinate transformation along the internal deformations. The quantum version of the period matrix can be written as 
\begin{align}
\tau_{IJ}=\oint_{\mathfrak{b}_I}\Omega_J=\frac{i}{2\pi}\frac{\partial\mathcal{A}_I}{\partial B_J},
\end{align}
with the same assumptions as above while taking the derivatives. The direct meaning of each entry can be also understood if we recast this equation as
\begin{align}
d\mathcal{A}_I=-2\pi i\sum_J \tau_{IJ}\,dB_J,
\end{align}
thus $\tau_{IJ}$ measures how the dual quantum period $\mathcal{A}_I$ changes when perturbative quantum periods $B_J$ change. A non-vanishing off-diagonal entry measures this response between different members of the chosen cycle basis. 

We can also interpret the period matrix $\tau$ as a relation for the formal Voros multipliers $\mathcal{X}$,
\begin{align}
\mathcal{X}_{\mathfrak{b}_I}=\exp\left( i\oint_{\mathfrak{b}_I}\lambda\right),
\end{align}
then we have
\begin{align}
\frac{\partial}{\partial B_J}\log \mathcal{X}_{\mathfrak{b}_I}=2\pi i\,\tau_{IJ}.
\end{align}

In the coordinates $(B_1,B_2,\mathcal{R}_q,a_3,a_4)$, the flow $\widehat{\mathcal{V}}$ keeps both quantum actions $B_1,B_2$ fixed, and we can show that the flow commutes with the derivatives with respect to the perturbative quantum periods,
\begin{align}
\left[ \widehat{\mathcal{V}},\partial_{B_J}\right]=0.
\end{align}
We can act with on both sides of Eq. (\ref{sexticPNP}) with $\partial_{B_J}$, we get a flow equation for the quantum corrected period matrix $\tau$,
\begin{align}
\widehat{\mathcal{V}}\tau_{IJ}=\frac{i}{16\pi g^2}\left[ (a_4-2)\frac{\partial^2 a_0}{\partial B_I\partial B_J}+3a_3\frac{\partial^2 a_1}{\partial B_I\partial B_J}\right],\qquad I,J=1,2.
\end{align}
We can equivalently express the same flow equation in terms of the P/NP Hamiltonian as
\begin{align}
\widehat{\mathcal{V}}\tau_{IJ}=\frac{i}{2\pi} \frac{\partial^2 \mathscr{H}}{\partial B_I\partial B_J},
\end{align}
The full quantum relation between perturbative and dual periods evolves while the perturbative quantum periods remain fixed. At leading order it becomes an evolution of the ordinary complex structure.


\subsection{P/NP Hamiltonian}\label{hamiltonian}
In this section, we would like to provide a geometrical picture for the genus-two P/NP relation and then argue its integrable structure by expressing it as an equality of differential one-forms instead of the matrix equation we just derived. Away from $g=0$, the extended coefficient space $\mathcal{M}$ is six dimensional, 
\begin{align}
\mathcal{M}=\{(g,a_0,a_1,a_2,a_3,a_4) \}.
\end{align}
We can equivalently replace the coefficient $a_2$ with the quantum residue $\mathcal{R}$ as a coordinate on $\mathcal{M}$, since the fixed residue depends on $g$, $a_2$ and $a_4$. As we elaborated before the coefficients $a_0$ and $a_1$ are distinguished from the rest of other coefficients because they deform the curve in such a way that the induced variation of $\lambda$ is holomorphic. Accordingly, we define an external parameter space $\mathcal{B}_{ext}=\{\eta\equiv(g,\mathcal{R},a_3,a_4)\}$ by collecting the parameters to change the singularity structure of the WKB differential. We can look at the whole extended coefficient space $\mathcal{M}$ as a fibration; $\pi:\mathcal{M}\rightarrow\mathcal{B}_{ext}$, such that we can use $(a_0,a_1)$ or $(B_1,B_2)$ as coordinates along the fiber $\mathcal{F}_\eta=\pi^{-1}(\eta)$, assuming the Jacobian has a non-vanishing determinant (as we will assume in this note). On a fixed fiber, we have $dg=d\mathcal{R}=da_3=da_4=0$, and only $da_0$ and $da_1$ are non-vanishing. The polar data of the WKB differential at infinity remain fixed as we move along the fiber. A vertical tangent vector has the form
\begin{align}
\mathcal{Y}=y_0\partial_{a_0}+y_1\partial_{a_1},
\end{align}
which will definitely change the perturbative periods when acting on them. These parameters are energy/source-like variables, as we vary them they scan a two-parameter family of systems with the same external asymptotic data, such as coupling, asymptotic monodromy and polar data. This is not a contradiction with how we defined the Whitham flow, which keeps the scaled quantum residue and perturbative periods fixed. The Whitham flow is in fact a horizontal flow from one fiber to another one, as its action on the external parameters is non-vanishing; for example $\mathcal{V}g=g$ and $\mathcal{V}a_4=c_4$. The Whitham deformation changes the external data, but retunes $a_0$, $a_1$ such that the perturbative quantum periods $B_1$ and $B_2$ remain fixed. 

Let us define the one-form $\Theta$ by
\begin{align}
\Theta& \equiv\sum_{I=1}^{2}\mathcal{F}_I\,dB_I,\\
&=\mathcal{F}^{T}dB,
\end{align}
taking $\mathcal{F}_i$ as components of a column matrix. On the hypersurface defined by fixed $(g,\mathcal{R},a_3,a_4)$, we can vary the holomorphic deformation parameters infinitesimally, which in turn changes the perturbative periods 
\begin{align}
dB=\mathcal{J}\,da,
\end{align}
leading to 
\begin{align}
\Theta&=\mathcal{F}^{T}\mathcal{J}da,\\
&=-\mathcal{I}^{T}da,
\end{align}
where we made use of the genus-two P/NP relation. Equivalently, we write
\begin{align}
\Theta=\sum_{I=1}^{2}\left( \widehat{\mathcal{V}}\mathcal{A}_I\right)dB_I&=\frac{1}{8g^2}(a_4-2)\,da_0+\frac{1}{8g^2}3a_3\,da_1,\\
&=d_{fib}\left(\frac{1}{8g^2}(a_4-2)\,a_0+\frac{1}{8g^2}3a_3\,a_1 \right),
\end{align}
showing that $\Theta$ is in fact exact along the fiber direction; $(a_3,a_4)$ are kept fixed and we vary $(a_0,a_1)$. This motivates us to define the Hamiltonian $\mathscr{H}$ for the P/NP relation by
\begin{align}
\mathscr{H}\equiv\frac{1}{8g^2}\left[(a_4-2)\,a_0+3a_3\,a_1 \right]
\end{align}
We can now write 
\begin{align}
\Theta=\frac{1}{8g^2}\,d_{fib}\left[(a_4-2)\,a_0+3a_3\,a_1 \right],
\end{align}
where we have been using the holomorphic deformation parameters as the coordinates along the fiber. As we emphasized before, we can equally use the quantum periods as coordinates, then in them the exterior derivative $d_B$ along the fiber direction becomes
\begin{align}
\Theta=d_B \mathscr{H}=\sum_{I=1}^{2}\left. \frac{\partial \mathscr{H}}{\partial B_I}\right|_{B_{J\neq I},\mathcal{R}_q,a_3,a_4}dB_I.
\end{align}
Using the definition of $\Theta$, we find an alternative form of the genus-two P/NP relation:
\begin{align}
\boxed{\widehat{\mathcal{V}}\mathcal{A}_I= \left. \frac{\partial \mathscr{H}}{\partial B_I}\right|_{B_{J\neq I},\mathcal{R}_q,a_3,a_4} }
\end{align}

Note that $\widehat{\mathcal{V}}$ differentiates along the Whitham flow, changing external parameters while keeping $B_I$ fixed, whereas $d_B$ differentiates within a fixed-external-parameter fiber, changing $B_I$. In fact, the P/NP rerelation connects these two transverse operations: Whitham derivative of the dual periods are given by the $B$-gradient of the scalar Hamilton function.


\subsection{Elliptic reduction of the genus-two P/NP relation}
A genus-two WKB curve associated with a quintic or sextic potential admits a degree-two elliptic quotient only on special loci in coefficient space. In the polynomial coordinate $z$, three cases arise: a sextic with reflection symmetry, a sextic with reciprocal symmetry and a quintic with reciprocal symmetry. The classical WKB differential $p\,dz$ descends directly to the quotient when the sextic potential is even about a point chosen as the origin and the involution acts as $(p,z)\mapsto(-p,-z)$. This simultaneous reversal of position and momentum preserves $p\,dz$. As argued in \cite{Fischbach:2018yiu}, the same involution preserves each higher-order WKB differential, so the descent holds order by order in the WKB expansion. We test our genus-two P/NP relation by specializing it to a general symmetric triple-well potential first, and the use the potential $V(z)=T_3(z)^2$, where $T_3(z)=4z^3-3z$ is the third Chebyshev polynomial, as an example. We show that the reduced genus-one P/NP relation is equivalent to the genus-one quantum Wronskian relation shown in \cite{Basar:2017hpr}.  

Although the symmetric family of sextic curves does not have a linear term, we include it anyways in our derivation because it is one of the holomorphic directions; we will set its coefficient to zero after we take the necessary derivative. We will write the Schr\"{o}dinger equation and the curve as
\begin{align}
g^2\psi''+Q(z)\psi=0,\qquad Q(z)=u-z^6-az^4-bz^2-vz, \qquad \Sigma: \,p^2=Q(z),
\end{align} 
where we redefined the coefficients to simplify the notation. The WKB one-form is given by
\begin{align}
\lambda_\Sigma=\frac{p}{g}dz.
\end{align}
The genus-two P/NP relations reduce to the following equations, 
\begin{align}
\left.\widehat{\mathcal{V}}\mathcal{A}_L\right|_{v=0}&=\frac{2-a}{8g^2}\left.\left(\frac{\partial u}{\partial B_L}\right)_{B_R,a,b,g}\right |_{v=0},\\
\left.\widehat{\mathcal{V}}\mathcal{A}_R\right|_{v=0}&=\frac{2-a}{8g^2}\left.\left(\frac{\partial u}{\partial B_R}\right)_{B_L,a,b,g}\right |_{v=0}.
\end{align}
By the symmetry of the potential when we set $v=0$, the perturbative and non-perturbative quantum periods on the left and right well become identical $B=\left. B_L\right|_{v=0}=\left. B_R\right|_{v=0}$ and $\mathcal{A}=\left. \mathcal{A}_L\right|_{v=0}=\left. \mathcal{A}_R\right|_{v=0}$, respectively, such that we have only one equation describing the flow of the non-perturbative quantum period $\mathcal{A}$ in terms of the perturbative quantum period. 

Next, we show that the reduced equation is indeed the genus-one P/NP relation of the elliptic quotient $\mathcal{E}$ of the sextic WKB curve $\Sigma$. The involution $ \iota:\,(p,z)\mapsto(-p,-z)$ keeps the curve and the WKB one-form invariant. Let us choose the invariant coordinates $x=z^2$ $q=zp$ , then we can write the curve $\mathcal{E}$ as 
\begin{align}
\mathcal{E}:\, q^2=x(u-x^3-ax^2-bx), \qquad \lambda_{\mathcal{E}}=\frac{q}{2gx}\,dx.
\end{align}
If we define the map $\pi(p,z)=(zp,z^2)$, then the pullback $\pi^{*}\lambda_{\mathcal{E}}=\lambda_\Sigma$, for the full quantum differentials of the inherited Schr\"{o}dinger operator, the same identity holds order by order. 

We follow the same recipe to construct the P/NP relation for the elliptic quotient $\mathcal{E}$, and then obtain the same expression as we did before by reducing the genus-two relation:
\begin{align}
\widehat{\mathcal{V}}\mathcal{A}=\frac{2-a}{16g^2}\left(\frac{\partial u}{\partial B}\right)_{a,b,g}.
\end{align}

However, it is known that the canonical involution $(z, y) \mapsto (-z,-y)$ fixes the action differential and by the Riccati recursion each even WKB differential \cite{Fischbach:2018yiu}. The corresponding quotient has elliptic equation $w^2 = 2x\left(E - x(4x- 3)^2\right)$, where $(x,w) = (z^2, z y)$. In particular, reflection symmetry exchanges the left and right outer wells and their adjacent barriers, so the corresponding well quantum periods satisfy $B_L=B_R\equiv B$ and the dual periods satisfy $\mathcal{A}_L=\mathcal{A}_R\equiv\mathcal{A}$.

We want to show how our genus-two P/NP equation reduces to a genus-one relation and reproduces the result of \cite{Basar:2017hpr}. For a comparison, let us rewrite the Schr\"{o}dinger equation with the Chebyshev polynomial in our normalization:
\begin{align}
\left[-\frac{g^2}{2} \partial_z^2+\frac{1}{2}z^2\left( z^2-\frac{3}{4}\right)^2 \right]\psi =\xi\psi.
\end{align}

The residue at infinity $\mu=0$, and we choose $\mathcal{V}=g\partial_g$, such that the velocity is fixed by $\mathcal{V}g=g$. Although $a_1$, one of the holomorphic coordinates is vanishing, we keep $a_1$ independent until both well periods have been fixed, and then we will set its value. We can evaluate the Jacobian,
\begin{align}
\mathcal{J}=\frac{\partial(B_L,B_R)}{\partial(a_0,a_1)}=\begin{pmatrix}
   - \partial_u B& d \\
    -\partial_u B & -d
  \end{pmatrix},
\end{align}
where we used the symmetry of the potential to equate the perturbative quantum periods on both left and right wells\footnote{The potential in question is a triple well, in addition to the left and right wells there is also a middle well. We can show that perturbative periods add up to the residue at infinity up to a sign. Combined with the parity symmetry of the well, the period of the middle well is twice of the left/right well, our choice does not affect the lift.} and $a_0=-u=2gE$. Furthermore, we defined $d=\partial_{a_1}B$. On a regular period patch the Jacobian $\mathcal{J}$ is invertible. We need to impose that both perturbative periods must be kept invariant under the lifted flow $\widehat{\mathcal{V}}$,
\begin{align}
\widehat{\mathcal{V}}=\left.g\partial_g\right|_{u}-\frac{g(\partial_g B)_u}{(\partial_u B)_g}\partial_u,
\end{align}
where we use the subscripts $u$ and $g$ indicates that these parameters must be kept fixed during differentiation. Specializing of Eq. (\ref{generalPNP}) for the Chebyshev potential, we get
\begin{align}
\begin{pmatrix}
   - \partial_u B& -\partial_u B\\
    d & -d
  \end{pmatrix}\begin{pmatrix}
   \widehat{\mathcal{V}}\mathcal{A}\\
     \widehat{\mathcal{V}}\mathcal{A}
  \end{pmatrix}=\begin{pmatrix}
-3/(16g^2)\\
0
  \end{pmatrix},
\end{align}
which reduces to a single equation,
\begin{align}
(\partial_u B)(\widehat{\mathcal{V}}\mathcal{A})=\frac{3}{32g^2}.
\end{align}
We can write the associated P/NP equation in a more familiar form as 
\begin{align}\label{ChebyshevPNP}
\left.g\frac{\partial \mathcal{A}}{\partial g}\right|_B=\left. \frac{3}{16g}\frac{\partial E}{\partial B}\right|_g
\end{align}

One may argue the fact that the pair of P/NP relations reducing to one does not necessarily imply the resulting relation to be a genus-one P/NP relation. We therefore would like to rederive the same relation starting with the elliptic quotient of the sextic curve and then apply our method. Let us consider the map $\pi$ that sends a point on the sextic curve to a point on the elliptic curve, $\pi:\,\Sigma\rightarrow\mathcal{E}$, such that $(z,y)\mapsto(x^2,xy)$

Moreover, we want to show that we can directly relate our final P/NP relation Eq. (\ref{ChebyshevPNP}) to the relation given in Eq. (4.1) of \cite{Basar:2017hpr} after matching the notation $(\hbar,s,a(s,\hbar),a^D(s,\hbar))$ used in that reference to ours $(g,E,B,\mathcal{A})$:
\begin{align}
\hbar=4\sqrt{2}g,\qquad s=32gE,\qquad B=\frac{a(s,\hbar)}{2\pi\hbar},\qquad \mathcal{A}=\frac{a^D(s,\hbar)}{i\hbar}.
\end{align} 
Let us start with rewriting the righthand side of our P/NP relation in the notation of  \cite{Basar:2017hpr}:
\begin{align}\nonumber
\left. \frac{3}{16g}\frac{\partial E}{\partial B}\right|_g&=\frac{3}{16g}\frac{1}{32g}\left.\frac{\partial s}{\partial B}\right|_\hbar=\frac{3}{16g}\frac{1}{32g}\frac{2\pi\hbar}{(\partial a/\partial s)_\hbar}\\
&=\frac{3\pi}{8\hbar\,(\partial a/\partial s)_\hbar}.
\end{align}
 
The left hand side requires slightly more work to rewrite: note that although $\widehat{\mathcal{V}}B=0$, the action of $\widehat{\mathcal{V}}$ on $a(s,\hbar)$ is non-trivial because $\widehat{\mathcal{V}}\hbar=\hbar$,
\begin{align}
\widehat{\mathcal{V}}a(s,\hbar)=a(s,\hbar).
\end{align}
On the other hand, we can evaluate the same differentiation using the chain rule and find
\begin{align}
\widehat{\mathcal{V}}s=\frac{a-\hbar (\partial a/\partial \hbar)_s}{(\partial a/\partial s)_\hbar}.
\end{align} 
It is also not hard to show that 
\begin{align}
i\hbar\widehat{\mathcal{V}}\mathcal{A}=(\partial a^D/\partial s)_\hbar \,\widehat{\mathcal{V}}s-\left( a^D-\hbar(\partial a^D/\partial \hbar)_s\right).
\end{align}
Once we put all of these differentiations together, we recover the Eq. (4.1) of \cite{Basar:2017hpr} 
\begin{align}
\left[a-\hbar\left(\frac{\partial a}{\partial \hbar}\right)_s \right]\left(\frac{\partial a^D}{\partial s}\right)_\hbar-\left[a^D-\hbar\left(\frac{\partial a^D}{\partial \hbar}\right)_s \right]\left(\frac{\partial a}{\partial s}\right)_\hbar=\frac{3\pi i}{8}.
\end{align}


\section{Octic Potential}\label{octic}

The derivation of the P/NP relations for the octic potential is very similar to the one of the sextic, hence we will be very brief in this section. We choose the potential to be 
\begin{align}
g^2\psi''+Q(z)\psi=0, \mbox{with}\qquad Q(z)=-(z^8+\sum_{j=0}^{6}a_j z^j),
\end{align}
where $a_0$ can be identified with the scaled energy $\xi$. We start again with the residue of the potential
\begin{align}
\mathcal{R}_{\pm}=\operatorname{Res}_{\pm\infty}\left( \frac{p}{g}dz\right)=\mp\frac{i}{2g}\left( a_3-\frac{1}{2}a_5a_6\right)\equiv \mp\frac{i}{2g}\mu,
\end{align}
where we changed the definition of $\mu$ while switching from the sextic potential to the octic. As it was the case for the sextic, the residue of the octic potential does not receive any quantum corrections because they behave as $p_{2n}(z)=\mathcal{O}(z^{4-10n})$, not producing any contributions to the residue. The most general external vector field preserving the quantum residue $\mathcal{R}$ is 
\begin{align}
\mathcal{V}_{\rm gen}=\sigma g\partial_g+c_6\partial_{a_6}+c_5\partial_{a_5}+c_4\partial_{a_4}+\left[ \sigma\mu+\frac{1}{2}\left( a_5c_6+a_6c_5\right)\right]\partial_{a_3},
\end{align}
where we used for the rates of change of the external parameters $\sigma=\dot{g}/g$ and $c_j=\dot{a}_j$. On a fixed residue $\mathcal{R}$ leaf we chose $(g,a_4,a_5,a_6)$ as our external parameters, since $a_3$ is determined by the other parameters on this leaf. We can find four external fields to generate P/NP relations for the octic potential:
\begin{align}\label{octicchoices}
\mathcal{V}_g&=g\partial_g+\mu\partial_{a_3},\\
\mathcal{V}_4&=\partial_{a_4},\\
\mathcal{V}_5&=\partial_{a_5}+\frac{1}{2}a_6\partial_{a_3},\\
\mathcal{V}_6&=\partial_{a_6}+\frac{1}{2}a_5\partial_{a_3}
\end{align}

The residue involves $(g, a_3, a_5, a_6)$. Let us hold $a_4$ fixed and choose the logarithmic coupling rate, then the residue preserving vector field reduces to
\begin{align}
\mathcal{V}= g\partial_g+c_6\partial_{a_6}+c_5\partial_{a_5}+\left[ \mu+\frac{1}{2}\left( a_5c_6+a_6c_5\right)\right]\partial_{a_3},
\end{align}

In general, all of these fields change the perturbative quantum periods. We need to lift these vector fields along the internal directions spanned by $(a_0, a_1, a_2)$ to make the periods $(B_1, B_2, B_3)$ invariant under the flows. The lifted field $\widehat{\mathcal{V}}$ can be written in terms of the holomorphic-period Jacobian as
\begin{align}
\widehat{\mathcal{V}}=\mathcal{V}-\sum_{\alpha=0}^{2}\sum_{I=1}^{3}\left(\mathcal{J}^{-1}\right)^{\alpha I}b_I\partial_{a_{\alpha}},
\end{align}
where we have defined $b_I=\mathcal{V}B_I$. For the minimal choice, we can set $c_6 = c_5 = 0$ and insert $b_I = g\partial_g B_I + \mu\partial_{a_3}B_I$ in the vector field $\widehat{\mathcal{V}}$ above. 

The last step to get the P/NP relation requires us to determine the bilinear pairing residue; it is a straightforward computation to compute it. If we store the rate of change of the coupling constant $\sigma$, the general P/NP relation for the octic potential takes the following form:
\begin{align}
\boxed{\widehat{\mathcal{V}}\mathcal{A}_I=\frac{1}{g^2}\left[ \frac{2\sigma a_6-5c_6}{30}\left(\frac{\partial a_0}{\partial B_I} -2a_6\frac{\partial a_2}{\partial B_I}\right)+\frac{3\sigma a_5-5c_5}{20}\frac{\partial a_1}{\partial B_I}+\frac{4\sigma a_4-5c_4}{10}\frac{\partial a_2}{\partial B_I}\right],}
\end{align}
for $I=1,2,3$. As we previously mentioned, the choice for the residue preserving vector field is not unique, and by lifting different choices among them correspond to different P/NP flows. We list them for the vector field presented in Eq. (\ref{octicchoices}):
\begin{align}\label{octicpnps}
g^2\widehat{\mathcal{V}}_g\mathcal{A}_I&=\frac{a_6}{15}\left( \frac{\partial a_0}{\partial B_I}-2a_6 \frac{\partial a_2}{\partial B_I}\right)+\frac{3a_5}{20}\frac{\partial a_1}{\partial B_I}+\frac{2a_4}{5}\frac{\partial a_2}{\partial B_I},\\
g^2\widehat{\mathcal{V}}_4\mathcal{A}_I&=-\frac{1}{2}\frac{\partial a_2}{\partial B_I},\\
g^2\widehat{\mathcal{V}}_5\mathcal{A}_I&=-\frac{1}{2}\frac{\partial a_1}{\partial B_I},\\
g^2\widehat{\mathcal{V}}_6\mathcal{A}_I&=-\frac{1}{6}\left(\frac{\partial a_0}{\partial B_I}-2a_6\frac{\partial a_2}{\partial B_I} \right).
\end{align}


\section{Quantum Corrected Sextic Potential}\label{qcsextic}

We would like to demonstrate that our approach is capable to produce the associated P/NP relation for quantum corrected potentials for higher genus potentials as well. As an example we choose the following potential,
\begin{align}
\left[ -g^2 \partial_z^2+W(z)^2 +\zeta g W'(z)+v z\right]\psi=\xi \psi, 
\end{align}
where we specifically choose the ``superpotential" $W(z)$ to be of the following form
\begin{align}
W(z)=z^3-az+b.
\end{align}
We will keep the parameters $a$, $b$ and $\zeta$ fixed along the flow, and choose as our holomorphic coordinates $v$ and $\xi$, similar to the classical sextic curve we earlier studied. We restrict the classical part of the potential $W(z)^2$ (with $v=0$) to give rise to a triple well potential with three real minima, which coincide with the real zeros of $W(z)$ if $4a^3>27b^2$ with $a>0$. Let us denote the zeros by $r_L<r_C<r_R$. Remember that the WKB curve is determined by the classical part of the potential; however, the quantum correction to the potential modifies the Riccati equation and therefore the WKB differential.

Our deformed quantum corrected potential can be generically written as 
\begin{align}
U(z;g,\eta)=V_0(z;\eta)+gV_1(z;\eta),
\end{align}
where $\eta$ is a collective label for the set of the deformation parameters. For completeness, let us briefly review the modifications to the Riccati equation and the quantum corrections induced by it. Suppose the wave function $\psi(z)$ has the form of the WKB Ansatz 
\begin{align}
\psi(z)=\mathcal{N}\exp\left[ \frac{i}{g}\int_{z_0}^{z}p(z')dz'\right],
\end{align}
which leads to the Riccati equation upon substituting into the Schr\"{o}dinger equation,
\begin{align}
p^2-igp'=2(\xi-U).
\end{align}
We can formally solve for $\rho$ as a formal expansion in $g$ by taking
\begin{align}
p^{\pm}(z;g,\eta)=\sum_{n\geq0}g^np^{\pm}_{n}(z;g,\eta).
\end{align}
At the leading order, we start with the classical momentum on formal branches
\begin{align}
p^{\pm}=\pm p_0=\pm\sqrt{2(\xi-V_0)}.
\end{align}
Away from the turning points, we can obtain the coefficients in the expansion recursively by solving the following equation,
\begin{align}\label{Riccati}
p^{\pm}_n=\frac{1}{2p^{\pm}_0}\left [ i\left (p^{\pm}_{n-1} \right)'-\sum_{j=1}^{n-1}p^{\pm}_jp^{\pm}_{n-j}-2V_n \delta_{n,1}\right].
\end{align}
The crucial difference from the classical potential with no quantum correction is that the first correction is \textit{not} a total derivative; for example we get
\begin{align}
p^{+}_1=\frac{i}{2}\left[\frac{d}{dz}\log (p_0)\right]-\frac{V_1}{p_0}.
\end{align}
Although the quantum corrections to the potential spoils the odd-even splitting of the WKB quantum corrections to the WKB differential, we can still define branch-antisymmetric momentum by
\begin{align}
\mathfrak{p}\equiv\frac{1}{2}\left( p^{+}-p^{-}\right),
\end{align}
and can further show that the symmetric combination of the correction is a total derivative,
\begin{align}
p^{+}+p^{-}=ig\frac{d}{dz}\log \mathfrak{p},
\end{align}
allowing us to express the WKB wave function only in terms of the branch-antisymmetric momentum,
\begin{align}
\psi^{\pm}(z)\frac{\mathcal{N}_{\pm}}{\sqrt{\mathfrak{p}(z)}}\exp\left[ \pm i \int_{z_0}^{z}\mathfrak{p}(z')dz' \right].
\end{align}
We will use the branch-antisymmetric momentum $P$ to define the WKB one-form and evaluate the quantum periods and residues,
\begin{align}
\lambda=\frac{1}{g}\mathfrak{p}\,dz.
\end{align}
The first step in the derivation of the P/NP relation is the determination of the vector field keeping the residue invariant. The WKB differential $\lambda$ has poles at infinity, 
\begin{align}
\mathcal{R}_{\pm}\equiv\operatorname{Res}_{\pm\infty} \lambda=\mp\frac{3i\zeta}{2}.
\end{align}
The quantum corrections from the Riccati equation begin at $x^{-5}$, become more singular with increasing degree and cannot change the residue from the zeroth order. The vector field $\mathcal{V}=g\partial_g$ keeps the residue fixed since it does not depend on $g$. Note that the internal classical variations with respect to two holomorphic directions $\xi$ and $v$ are given by
\begin{align}
\partial_\xi\lambda_0=\frac{dz}{2gp_0},\qquad \partial_v\lambda_0=-\frac{z\,dz}{2gp_0},
\end{align}
and the residue remains invariant in these directions as well, $\partial_\xi \mathcal{R}_{\pm}=\partial_v \mathcal{R}_{\pm}=0$. 

In general, the residue preserving vector field $\mathcal{V}$ does not keep the perturbative quantum periods $B_{L,R}$ fixed and we need to construct its horizontal lift $\widehat{\mathcal{V}}$ keeping using the internal holomorphic coordinates,  $\xi$ and $v$ for the present quantum corrected potential, 
\begin{align}
\widehat{\mathcal{V}}=\mathcal{V}+\zeta^{\xi}\partial_{\xi}+\zeta^{v}\partial_v. 
\end{align}
We can determine these components easily using the holomorphic-period Jacobian $\mathcal J$,
\begin{align}
\mathcal{J}= \begin{pmatrix}
    \partial_\xi B_L & \partial_v B_L \\
    \partial_\xi B_R & \partial_v B_R 
  \end{pmatrix},
\end{align}
and they are given by 
\begin{align}
 \begin{pmatrix}
\ell^{\xi}  \\
\ell^{v}
  \end{pmatrix}= -\mathcal{J}^{-1}\mathcal{V}\begin{pmatrix}
     B_L  \\
    B_R 
  \end{pmatrix}.
\end{align}
Following our general recipe the final step consists of employing the Riemann bilinear identity with the appropriate one-forms and primitive:
\begin{align}
\Omega=\widehat{\mathcal{V}}\lambda,\qquad d\Phi=\Omega,\qquad \omega_\xi=\partial_\xi\lambda,\qquad\omega_v=\partial_v\lambda.
\end{align}
The one-forms are chosen to make half of the summand appearing in the Riemann bilinear identity to vanish giving us flow of the dual periods in a mixed form 
\begin{align}
\sum_{I=L,R}\mathcal{J}_{I\alpha}\,\widehat{\mathcal{V}}\mathcal{A}_I=-\sum_{P}\operatorname{Res}_P\left(\Phi\omega_\alpha\right).
\end{align}

The final step requires the evaluation the residue on the right hand side including the derivative quantum corrections coming from the WKB expansion. We note that it is crucial to take into account all simple poles in this computation, not just merely the contributions from the poles of the WKB differential, as the turning points may give rise to simple poles in the quantum corrections. As we will argue later, the bilinear residue is equal to the classical residue at infinity. On the other hand, the poles at the turning points do not contribute to it because there are no quantum corrections but they perfectly cancel each other. Let us expand the primitive and the one-forms in the residue using the following expansion. Set $x=1/z$, then we have
\begin{align}
\Phi&=\frac{i}{g}\left[-\frac{1}{4}x^{-4}+\frac{a}{2}x^{-2}+bx^{-1}\right]+\mathcal{O}(x),\\
\omega_{\xi}&=\frac{i}{2g}\left[x+ax^3+bx^4+\mathcal{O}(x^5)\right]dx,\\
\omega_{v}&=-\frac{i}{2g}\left[1+ax^2+bx^3+\mathcal{O}(x^4) \right]dx,
\end{align}
where the primitive $\Phi$ does not have a logarithmic term because $\mathcal{V}\rho_{\pm}=0$. Moreover, both factors in the residue change sign at the other infinity such that their product is equal on both poles. The P/NP equation takes the following form,
\begin{align}
J^{T}\widehat{\mathcal{V}}\begin{pmatrix}
     \mathcal{A}_L  \\
     \mathcal{A}_R 
  \end{pmatrix}=\frac{1}{4g^2}\begin{pmatrix}
     a \\
     -3b
  \end{pmatrix},
\end{align}
We should emphasize that since the quantum deformation parameter $\zeta$ does not explicitly appear in the P/NP equation, it has not been dropped. It also does not modify the particular bilinear pairing coefficients that we evaluated at infinities. On the other hand, it modifies the quantum residue and enters the period calculation. 

We can rewrite the P/NP relation as 
\begin{align}
\boxed{\left.g\frac{\partial\mathcal{A}_I}{\partial g}\right|_{B_L,B_R,a,b,\zeta}=\frac{1}{4g^2}\left[a\left.\frac{\partial \xi}{\partial B_I}\right|_{B_{J\neq I},g,a,b,\zeta}-3b\left.\frac{\partial v}{\partial B_I} \right|_{B_{J\neq I},g,a,b,\zeta}\right],\qquad I=L,R.}
\end{align}

We have checked the relation to the order $\mathcal{O}(g^2)$.

\section{Rational Potentials}\label{rationalpot}

\par{In this section, we construct a coupled genus-two P/NP relation for a smooth asymmetric rational double-well family. We would like to present this example as a test of the range of validity of our method. Rational potentials differ in a number of ways from our previous examples of polynomial potentials. For the sextic potential the only singularity was at infinity, which does not get any quantum corrections. Finite singularities arise only from turning points. The rational potentials have fixed finite poles and also contribute to the WKB geometry. Another distinctive feature of our rational potential is that the residue depends explicitly on the energy. The energy is not one of the holomorphic directions unlike the other examples we studied so far when applying our approach. Moreover, the WKB differential is modified. }

Our rational potential is 
\begin{align}
V(z)=z^2+\frac{\alpha+\beta z}{z^2+\gamma^2}, \qquad \gamma>0,
\end{align}
where we assume $\alpha^2+\beta^2+\gamma^2\neq0$ to keep the WKB curve smooth. For physical potential with real $\alpha$ and $\beta$, this is automatically satisfied as long as we insist on a rational potential. 
\begin{figure}[h]
  \begin{center}
    \includegraphics[width=0.4\textwidth]{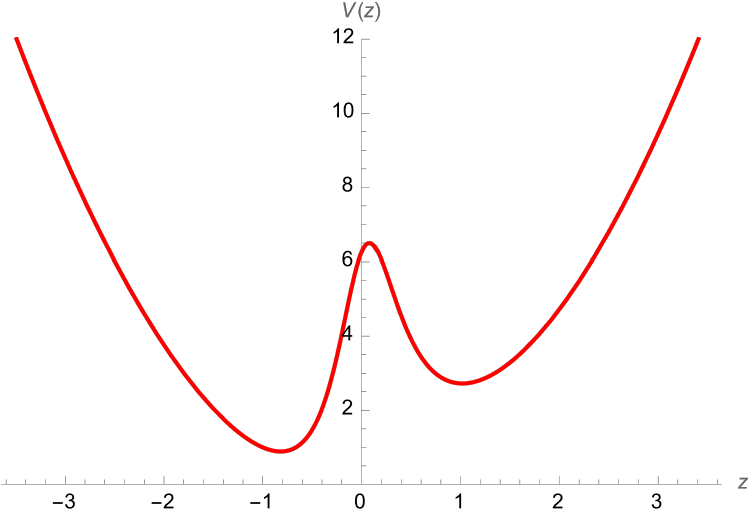}
  \end{center}
  \caption{The plot of our rational potential with $\alpha=\beta=1$ and $\gamma=0.4$.}\label{rational}
\end{figure}
For brevity, let us call the denominator $D=z^2+\gamma^2$. The classical momentum $p_0$ is given by 
\begin{align}
p_{0}^2=\xi-z^2-D^{-1}(\alpha +\beta z).
\end{align}
The WKB curve is a genus-two hyperelliptic curve given by
\begin{align}
y^2=D\left[ \left( \xi-z^2\right) D-\alpha-\beta z\right],
\end{align}
where we defined $y=Dp_0$. The quantum corrected WKB one-form is given by 
\begin{align}
\lambda_q=\frac{p}{g}dz=\frac{y}{gD}dz,
\end{align}
not just $(y/g)\, dz$ anymore. Note that the extra factor of $D$ in the denominator introduces two finite singularities to the WKB differential in addition to the one at infinity. The classical residue at $\pm\infty$ are given by
\begin{align}
\mathcal{R}_\pm\equiv\operatorname{Res}_{\pm\infty} \frac{p_0}{g}dz=\pm\frac{i\xi}{2g}.
\end{align}
which does not get any quantum correction by WKB expansion since $p_{2n}=\mathcal{O}(z^{1-4n})$ for $n\geq1$. As previously mentioned, there are two additional poles of $p_0$, where $D$ vanishes; but the WKB differential is regular at those points, and we can show that higher even order WKB differentials may have poles at those point but they do not generate any residue contribution. 

Following our prescription we first construct the vector field $\mathcal{V}$ that will keep the residue invariant: 
\begin{align}
\mathcal{V}=g\partial_g+\xi \partial_\xi, \qquad\mathcal{V}\mathcal{R}_\pm=0.
\end{align}
Next we need to lift the vector field $\mathcal{V}$ to keep the perturbative quantum periods fixed as well along the flow. $\alpha$ and $\beta$ parameters of the potential offer the holomorphic directions for the required lift since
\begin{align}
\partial_\alpha\lambda_0=-\frac{dz}{2gy}, \qquad \partial_\beta\lambda_0=-\frac{z\,dz}{2gy},
\end{align}
span the holomorphic differentials of the compact surface. The fixed-period lift $\widehat{\mathcal{V}}$ is given by 
\begin{align}
\widehat{\mathcal{V}}=\mathcal{V}+\zeta^{\alpha}\partial_\alpha+\zeta^{\beta}\partial_\beta, \qquad \widehat{\mathcal{V}}B_{L,R}=0.
\end{align}
Using the holomorphic-period Jacobian $\mathcal{J}$, we can solve for the coefficients for the lift
\begin{align}
 \begin{pmatrix}
\zeta^{\alpha}  \\
\zeta^{\beta}
  \end{pmatrix}= -\mathcal{J}^{-1}\mathcal{V}\begin{pmatrix}
     B_L  \\
    B_R 
  \end{pmatrix}.
\end{align}
By construction, the lifted flow satisfies $\widehat{\mathcal{V}}B_{L}=\widehat{\mathcal{V}}B_{R}=\widehat{\mathcal{V}}\rho_{\pm}=0$. Now, we are ready to derive the P/NP relation for the potential in question. Set 
\begin{align}
\Omega=\widehat{\mathcal{V}}\lambda_q, \qquad d\Phi=\Omega,\qquad \omega_\alpha=\partial_\alpha \lambda_q,\qquad \omega_\beta=\partial_\beta\lambda_q.
\end{align}
The Riemann bilinear identity takes the following form 
\begin{align}
\sum_{I=L,R}\left[\left( \oint_{\mathfrak{a}_I}\Omega \right) \left( \oint_{\mathfrak{b}_I}\omega_i \right) -\left( \oint_{\mathfrak{b}_I}\Omega \right) \left( \oint_{\mathfrak{a}_I}\omega_i \right) \right]=2\pi i\sum_P\operatorname{Res}_{P} \left(\Phi\omega_i \right),
\end{align}
where we mean $i=\alpha, \beta$. Note that the first term in the summand again vanishes by construction such the bilinear identity gives rise to the P/NP relation,
\begin{align}
\sum_{I=L,R}\left(\partial_i B_I \right)\widehat{\mathcal{V}}\mathcal{A}_I=-\sum_P\operatorname{Res}_{P} \left(\Phi\omega_i \right), \qquad i=\alpha,\,\beta.
\end{align}
Finally, we need to evaluate the bilinear pairing residue, which we sketch in Appendix \ref{cancellation}. Using the inverse of the perturbative Jacobian $\mathcal{J}$, we obtain the P/NP relation for our rational potential that we will test:
\begin{align}
\boxed{\widehat{\mathcal{V}}\mathcal{A}_I= \frac{1}{2g^2}\left.\frac{\partial \alpha}{\partial B_I}\right|_{B_{J\neq I},g,\mathcal{R}_{\pm},\ell }.}
\end{align}

\section{Non-hyperelliptic Curves}\label{nonhyper}

Third-order differential equations arise naturally in the quantization of Seiberg–Witten curves. An example relevant to our discussion is provided by the four-dimensional $\mathcal{N}=2$ Argyres–Douglas theories of type $(A_2,A_N)$. On the family $y^3+q(z)=0$, the classical momentum $y$ has three branches and the Seiberg–Witten differential is $y\,dz$. In the Nekrasov–Shatashvili limit, $y$ is represented by $\hbar\partial_z$, so the cubic equation for the momentum gives the third-order equation $\bigl[\hbar^3\partial_z^3+q(z)\bigr]\psi=0$. The same equation appears in the conformal limit of the linear problem for $A_2^{(1)}$ affine Toda theory, where the correspondence between ordinary differential equations and integrable models relates its WKB periods to thermodynamic Bethe Ansatz equations. For the $(A_2,A_3)$ theory, $q(z)$ is quartic and the smooth compactified curve has genus three. This gives us a setting in which to study the P/NP relation directly in terms of quantum periods, beyond the usual interpretation in terms of wells and tunneling barriers.

We start with the third-order equation\footnote{We are keeping the notation of \cite{Ito:2021sjo}.}
\begin{align}
 \bigl[\hbar^3\partial_z^3+q(z)\bigr]\psi=0,\qquad
 q(z)=z^4+az^2+bz+c .
 \label{trigonal-undeformed}
\end{align}
Its classical curve, $y^3+q(z)=0$, is a smooth plane quartic for generic coefficients and hence is non-hyperelliptic of genus three. We can ask the same question as for the octic oscillator, now for a momentum with three branches. How does each of the three dual periods change when we vary $a$ and keep all three perturbative actions fixed?

At fixed $\hbar$, Eq.~\eqref{trigonal-undeformed} leaves only $b$ and $c$ available to compensate a change in $a$. To impose three independent period conditions we introduce one more parameter, $u$, and consider
\begin{align}
 \bigl[\hbar^3\partial_z^3+v\hbar\partial_z+q(z)\bigr]\psi=0,\qquad
 \Sigma:\quad y^3+vy+q(z)=0 .
 \label{trigonal-operator}
\end{align}
The choice of the term $v\hbar\partial_z$ follows from the geometry of $\Sigma$. With $\Delta=3y^2+v$, implicit differentiation at fixed $z$ gives
\begin{align}
 (\partial_c,\partial_b,\partial_v)(y\,dz)
   =-\frac{dz}{\Delta}(1,z,y),\qquad
 \partial_a(y\,dz)=-\frac{z^2\,dz}{\Delta}.
 \label{trigonal-holomorphic}
\end{align}
The first three forms are the standard basis of holomorphic differentials on a smooth plane quartic. The last has a double pole at the unique point at infinity. Thus $a$ provides the external deformation, while $v$ supplies precisely the missing holomorphic direction. The genus is already three at $v=0$. We now apply our three steps of the P/NP construction.

We start by choosing residue-preserving flow. Writing $\psi=\exp(\hbar^{-1}\int P\,dz)$ gives
$P^3+vP+q+3\hbar PP'+\hbar^2P''=0$, with leading term $P=y$. We separate the logarithmic amplitude $-\tfrac12d\log\Delta$ and discard exact differentials. We compute the quantum periods along the cycles with the intersection matrix given by $\mathfrak{a}_I\cdot\mathfrak{b}_J=\delta_{IJ}$. The residues of $\lambda$ vanish. At a simple finite branch value, Abel's identity for the Wronskian implies that the sum over the three sheets of each WKB correction beyond the amplitude is exact. The differential is regular on the third sheet, and its residue at the ramification point vanishes. At infinity, putting $z=t^{-3}$ gives $y\sim\kappa t^{-4}$ with $\kappa^3=-1$. The expansion of $y\,dz$ has no $dt/t$ term and the higher WKB corrections are regular there. Both $\mathcal{V}_a=\partial_a$ and $\mathcal{V}_\hbar=\hbar\partial_\hbar$ therefore preserve the quantum residues.

Next we lift the flow to keep the perturbative quantum periods fixed. We denote the holomorphic direction by $\boldsymbol u=(v,b,c)$ and the Jacobian again by $\mathcal{J}_{I\alpha}=\partial_{u_\alpha}B_I$, computed from the full quantum periods. The leading coefficient of $\mathcal{J}$ is proportional to the matrix of $\mathfrak{a}$-periods of the holomorphic forms in Eq. \eqref{trigonal-holomorphic}. It is invertible, such that the three actions determine $(v,b,c)$ locally at fixed $(a,\hbar)$. For $\mathcal{V}_a$ and $\mathcal{V}_\hbar$ we denote the lifts by $\widehat{\mathcal{V}}_a$ and $\widehat{\mathcal{V}}_\hbar$. In either case, the compensating change in the internal parameters follows from the perturbative periods and they satisfy by construction,
\begin{align}
\widehat{\mathcal{V}}_a B_I=0, \qquad\widehat{\mathcal{V}}_\hbar B_I=0,\quad I=1,2,3.
 \label{trigonal-lift}
\end{align}

In the last step, we determine the variation of the dual ``non-perturbative'' periods using the Riemann bilinear identity. Put $\Omega_a=\widehat{\mathcal{V}}_a\lambda$. Since its $\mathfrak{a}$-periods vanish by our construction, the bilinear identity gives
\begin{align}
 \sum_{I=1}^3\mathcal{J}_{I\alpha}\, \left(\widehat{\mathcal{V}}_a\mathcal{A}_I\right)
 =\sum_{P}\operatorname{Res}_P
       \left[\left(\int^p\partial_{u_j}\lambda\right)\Omega_a\right].
 \label{trigonal-bilinear}
\end{align}
At classical order only the $v$-variation contributes. Indeed, at infinity a primitive of $\partial_v(y\,dz)$ starts with $t/\kappa$, while $\partial_a(y\,dz)$ starts with $dt/(\kappa^2t^2)$. Their product has residue $\kappa^{-3}=-1$. The primitives for $b$ and $c$ start at orders $t^2$ and $t^5$, respectively, and do not contribute to the residue calculation. At higher WKB orders the finite ramification points also contribute. For example, for each $s$ with $3s^2+v=0$, their contribution to the first correction in the $(v,a)$ pairing sums to
\begin{align}
 \left.\frac{s}{12}\frac{d}{dz}
   \left(\sum_{q(r)=z}\frac{r}{q'(r)}\right)\right|_{z=2s^3}=0 .
\end{align}
The sum in parentheses vanishes by partial fractions. Through unscaled WKB order $\hbar^6$ the finite residues cancel in the sums for the external and internal pairings, and power counting excludes further residues at infinity. In the order $(v,b,c)$ the right-hand side of Eq.~\eqref{trigonal-bilinear} is therefore $(-\hbar^{-2},0,0)+O(\hbar^6)$.

Since only the $v$-entry survives to this order, inverting $J^T$ gives the $a$-flow in terms of $(\mathcal{J}^{-1})_{vI}=\partial v/\partial B_I$. To obtain the $\hbar$-flow, assign weights $3$ and $4$ to $z$ and $y$. Equation~\eqref{trigonal-operator} then gives weights $6$ and $7$ to $a$ and $\hbar$, and the scaled periods are invariant under the resulting dilation. Thus $7 \widehat{\mathcal{V}}_\hbar+6a \widehat{\mathcal{V}}_a$ annihilates every period. The two P/NP relations are
\begin{equation}
 \boxed{\begin{aligned}
 \left.\partial_a\mathcal A_I\right|_{\boldsymbol B,\hbar}
   &=-\frac{1}{\hbar^2}\frac{\partial v}{\partial B_I}+O(\hbar^7),\\
 \left.\hbar\partial_\hbar\mathcal A_I\right|_{\boldsymbol B,a}
   &=\frac{6a}{7\hbar^2}\frac{\partial v}{\partial B_I}+O(\hbar^7),
 \qquad I=1,2,3 .
 \end{aligned}}
 \label{eq:trigonal-pnp}
\end{equation}
On the right, $a,\hbar$ and the other two actions are held fixed. These relations determine the dual-period variations through order $\hbar^5$ in the scaled periods at generic moduli. To recover the result on the original curve, we take the derivatives in the full family and then set $v=0$. Although $v$ vanishes on that slice, its derivative with respect to an action can remain nonzero.

\section*{Acknowledgments}
We would like to thank Atakan \c{C}avu\c{s}o\u{g}lu, Ilmar Gahramanov, and Kemal Tezgin for our previous fruitful collaborations and for many illuminating discussions. We also thank Emil Hakan Leeb-Lundberg, Osman Teoman Turgut, and Umut Varolg\"{u}ne\c{s} for many inspiring discussions. The results of this work were presented at a seminar in May 2026 in the Department of Mathematics at \.{I}stanbul Technical University before the manuscript was submitted. CK would like to thank the members of the department for the invitation and for the enlightening discussions. The work of CK is supported by the Scientific and Technological Research Council of Turkey (T\"{U}B\.ITAK) under the Grant Numbers 124F428 and 225N333. We thank T\"{U}B\.ITAK for their support. This article is based upon work from COST Action 21109 CaLISTA and COST Action 22113 THEORY-CHALLENGES, supported by COST (European Cooperation in Science and Technology). We acknowledge the use of AI agent ChatGPT Astra (Open AI) to help us in period computations, derivations and text editing. All results are independently verified analytically or with the aid of codes written in Mathematica (Wolfram) by the authors.
\appendix

\section{Bilinear pairing residue quantum correction cancellation}\label{cancellation}
In many examples, we observe that the residual of the Riemann bilinear identity does not either receive any quantum corrections or there is a perfect cancellation. This observation motivated us to consider whether $\Phi\omega_{\alpha}$ can be expressed as an exact form or as a holomorphic 1-form. In either case, we relate the parameter derivatives $\partial_{a_{\alpha}}$ and $\widehat{\mathcal{V}}$ to the coordinate derivatives. Consider a coordinate change $z\mapsto \varphi(z)$, such that Schrödinger equation remains invariant. Let the solution of the Schrödinger equation transform as 
\begin{equation}\label{psitrnasformation}
    \tilde{\psi}(z)=A(z)(\psi \circ \varphi)(z)
\end{equation}
Differentiate $\tilde{\psi}$, and we obtain
\begin{equation}
    \tilde{\psi}''=A''\psi \circ \varphi+(2A'\varphi'+A\varphi'')\psi'\circ \varphi+A(\varphi')^2\psi''\circ \varphi,
\end{equation}
which implies $2A'\varphi'+A\varphi''=0$. Then, we can integrate and get $A=1/\sqrt{\varphi'}$. Let us denote $Q/g^2=F$,
\begin{equation}
    \tilde{\psi}''=\left (\frac{A''}{A}-(\varphi')^2F\circ\varphi \right )\tilde{\psi}\implies \tilde{F}=(\varphi')^2 F\circ\varphi-\frac{A''}{A}.
\end{equation}
Since $A'/A=-\varphi''/2\varphi'$, $A''/A=(A'/A)'+(A'/A)^2=-\frac12 \{\varphi,z\}$. Then, $F$ transforms as follows,
\begin{equation}\label{Ftransformation}
    \tilde{F}=(\varphi')^2F\circ\varphi+\frac12 \{\varphi,z\}.
\end{equation}
Consider $\varphi$ as an infinitesimal coordinate change, i.e., $\varphi(z)=z+\varepsilon v_{\delta}(z)$, where the variation under the infinitesimal coordinate change is defined by $\varepsilon\, \delta \Upsilon=\widetilde{\Upsilon}-\Upsilon$. Then, the Schwarzian derivative term becomes $\{\varphi,z\}=\varepsilon v_{\delta}'''$. Using Eq. \eqref{psitrnasformation} and Eq. \eqref{Ftransformation}, we can write for the variation of $F$ 
\begin{equation}
    \delta F=v_{\delta}F'+2v_{\delta}'F+\frac12F''',
\end{equation}
and
\begin{equation}
    \Phi_{\delta}=v_{\delta}K+\frac{i}{2}v_{\delta}'.
\end{equation}
Here, we used the notation
\begin{equation}
    K = \frac{p_q}{g}, \quad K^2-iK'=\frac{Q}{g^2}=F.
\end{equation}
Then, the linearization of the Ricatti equation gives
\begin{equation}
    (2K-i\partial_x)\delta K=\frac{\delta Q-2(\delta g/g)Q}{g^2}.
\end{equation}
We can expand $v_\delta$ as a power expansion of $g$,
\begin{equation}\label{expansionvdelta}
v_\delta=v_\delta^{(0)}+g^2v_\delta^{(2)}+g^4v_\delta^{(4)}+\mathellipsis,
\end{equation}
for $p_0^2=Q$,
\begin{equation}
\delta\bigg(\frac{p_0}{g}dz\bigg)= \frac{\delta Q - 2(\delta g/g)Q}{2p_0g}dz
\end{equation}
The derivative of the primitive of the form $v_\delta^{(0)}p_0/g$ is that,
\begin{equation}
d\bigg(\frac{v_\delta^{(0)}p_0}{g}\bigg)= \frac{2Q(v_\delta^{(0)})'+Q'v_\delta^{(0)}}{2p_0g}dz.
\end{equation}
Simplification of the equation above is 
\begin{equation}
2Q(v_\delta^{(0)})'+ Q'v_\delta^{(0)}= \delta Q - 2\frac{\delta g}{g}Q
\end{equation}
Then, taking direct differential can be written as follows
\begin{align}\nonumber
    (2K-i\partial_x)\Phi_\delta' &= 2v_\delta'(K^2-iK')+v_\delta(2KK'-iK'') + \frac{1}{2}v_\delta''',\\
& =   \frac{2Qv_\delta'+Q'v_\delta+(g^2/2)v_\delta'''}{g^2}
\end{align}
Here, the terms of $Kv_\delta''$ cancels because of the corrections $iv_\delta/2$.
\begin{equation}\label{twofifteen}
  2Qv_\delta'+Q'v_\delta+(g^2/2)v_\delta'''= \delta Q - 2\frac{\delta g}{g}Q,
\end{equation}
which is an order by order recursion, use \eqref{expansionvdelta} and obtain
\begin{align}
    2Q\left(v^{(0)}_\delta\right)'+Q'v^{(0)}_\delta &= \delta Q - 2\frac{\delta g}{g}Q,\\
    2Q\left(v^{(2n)}_\delta\right)'+Q'v^{(2n)}_\delta+\frac{1}{2}\left(v_\delta^{(2n-2)}\right)'''&= 0.
\end{align}
We now show that the coordinate variation constructed above is regular at a simple turning point. The dependence on $z$ and the deformation parameters is otherwise arbitrary in the course of this argument, and is independent of all regularity statements are to be understood coefficientwise in the formal parameter $g$.

Let $z_0$ be a simple zero of $Q$, and assume that the
parameter variations of $Q$ are holomorphic near $z_0$.
The position of the zero is allowed to vary. To show regularity, introduce the ramified coordinate
$z-z_0=\zeta^2$. For any function $f$ holomorphic in a neighborhood of $z_0$, the differential $\frac{f\,\mathrm dz}{2p_0}$ is holomorphic on the $\zeta$-disc and changes sign under the involution $\zeta \mapsto -\zeta$. Its holomorphic primitive, let us denote as $h$, normalized by $h(0)=0$, is odd.
Since $p_0$ has a simple zero in $\zeta$, $h/p_0$ is holomorphic and even. It therefore becomes to a holomorphic function of $z$, and satisfies \eqref{twofifteen}. The solution is unique: a homogeneous solution would make $p_0v$ constant, whereas $p_0v$ is an odd holomorphic
function.

The same construction applies at a fixed simple pole of
$Q$, provided the variations have at most a
simple pole. The differential $f dz/2p_0$ is again holomorphic. Now $p_0$ has a simple pole in $\zeta$, so $h/p_0$ vanishes to order at
least two in \(t\). Its descended function therefore
satisfies $v(z_0)=0$. Hence the coordinate variation fixes the poles.

Take two variations $\delta_1,\delta_2$,
the linearized Riccati equations read
\[
\begin{split}
    (2K- i\partial_z)\delta_2K
    =\delta_2\left(\frac{Q}{g^2}\right),\\
        (2K- i\partial_z)\delta_1K
    =\delta_1\left(\frac{Q}{g^2}\right).
\end{split}   
\]
Using this equation and the product rule gives
\begin{equation}
    \Phi_{\delta_1}\,\delta_2\lambda=\frac12 v_{\delta_1}
       \delta_2\left(\frac{Q}{g^2}\right)
        dz
    +\frac{i}{2}
       \mathrm d\!\left(v_{\delta_1}\delta_2K\right).
    \label{bilineardecomposition}
\end{equation}

At a simple turning point, both \(v_{\delta_1}\) and
\(\delta_2(Q/g^2)\) are holomorphic. At a fixed simple
pole, the zero of \(v_{\delta_1}\) cancels the possible
simple pole of \(\delta_2(Q/g^2)\). In either case, the
first term on the right-hand side is holomorphic. Hence,
\begin{equation}
    \operatorname{Res}_{z_0}
    \left(\Phi_{\delta_1}\,\delta_2(K dz)\right)=0
    \label{all orders cancellation}
\end{equation}
coefficientwise for all quantum corrections. There are no quantum corrections to the residue of the Riemann bilinear identity in polynomial potentials; therefore, the quantum corrections around the turning points must cancel. In the case of potentials with finite poles, like the doubly elliptic periodic potential (a genus-1 hyperelliptic curve), the contributions from quantum corrections at poles are non-zero and do not cancel by themselves; rather, they cancel with the contributions from quantum corrections at turning points, order by order. This cancelation has already been observed order by order calculation for the first few orders \cite{Kozcaz:2026isf}.

\section{Periods}\label{periods}
In order to check the P/NP relation derive the Picard-Fuchs equations (classical) and quantum corrections following \cite{Fischbach:2018yiu}. Then, we check the P/NP relations we found using the periods we explicitly calculate. Here, we list the periods we used in those checks as reference.

\subsection{Sextic Potential}

Define $(p')^2\equiv-(z^6+a_0)$, and $\eta\equiv(-a_0)^{1/6}$, and $\zeta\equiv e^{i\pi/3}$. Then the roots of $(p')^2$ are $z_k=\eta \zeta^k$. Along the radial segments $z=\eta \zeta^k r$ with $r\in [0,1]$, $p'=\eta^3\sqrt{1-r^6}$. Let $\gamma_j$ be the path that goes from $z_{j}$ to the origin and then goes to $z_{j+1}$ along the other sheet and returns from $z_{j+1}$ to $z_j$ using the opposite sheets that have been used when we go this far. Then all $\gamma_j$ are closed loops, by definiton and $\gamma_j \circ \gamma_{j+1}=1$ and $\gamma_j \circ \gamma_{j-1} = -1$. We have more than one choice of A-cycles and B-cycles, if we choose the convention $\mathfrak{a}_1=\gamma_0$, $\mathfrak{a}_2=\gamma_2$, $\mathfrak{b}_1=\gamma_1+\gamma_3$, and $\mathfrak{b}_2=\gamma_3$, which satisfy $\mathfrak{a}_i \circ \mathfrak{b}_j=\delta_{ij}$ and $\mathfrak{b}_i \circ \mathfrak{a}_j=-\delta_{ij}$, and $\mathfrak{a}_i \circ \mathfrak{a}_j=0$, and $\mathfrak{b}_i \circ \mathfrak{b}_j=0$. We find the classical periods in the following closed formulas,
\begin{align}
    \Pi_j=\frac13 \sum_{\mathbf{n}\ge 0}\frac{(-1/2)_{l_{\mathbf{n}}}}{\mathbf{n}!}\mathbf{a}^{\mathbf{n}} (-a_0)^{(m+4)/6-l_{\mathbf{n}}}B\left ( \frac{m+1}{6}, \frac32 - l_{\mathbf{n}} \right )\zeta^{j(m+1)}\left ( \zeta^{m+1}-1\right)
\end{align}
Here, $\mathbf{n}=(n_1,n_2,n_3,n_4)$, $l_{\mathbf{n}}=\sum_{k=1}^4n_k$, $m=\sum_{k=1}^4kn_k$, $\mathbf{n}!=n_1!n_2!n_3!n_4!$, $\mathbf{a}^{\mathbf{n}}=a_1^{n_1}a_2^{n_2}a_3^{n_3}a_4^{n_4}$ .We obtain the periods up to total order of $a_i$'s (for $i\ne0$) 2,up to $g^4$. 

We have obtained the classical periods as linear combination of $\Pi$'s
\begin{equation}
\begin{split}
    B_1=\frac{\Pi_0}{2\pi g}, \qquad B_2=\frac{\Pi_2}{2\pi g},\\
    \mathcal{A}_1=\frac{\Pi_1+\Pi_3}{ig}, \qquad \mathcal{A}_2=\frac{\Pi_3}{ig}.
    \end{split}
\end{equation}

We obtain the quantum corrected periods by using the quantum correction operators we have introduced, $B_i(g,a)=\sum_{k\ge0}g^{2k}\mathcal{D}_{2k}B_i$, $\mathcal{A}_i(g,a)=\sum_{k\ge0}g^{2k}\mathcal{D}_{2k}\mathcal{A}_i$, and expand as follows
\begingroup\makeatletter\def\f@size{10}\check@mathfonts
\begin{align*}
B_1(g,a) &= \frac{1}{g} \Bigg( -\frac{a_2^2}{64 \sqrt{3} a_0^{4/3} N_1} - \frac{i a_2^2}{64 a_0^{4/3} N_1} + \frac{i a_2 a_4 N_1}{12 \sqrt{3} \pi a_0} - \frac{a_2 a_4 N_1}{36 \pi a_0} + \frac{a_1 a_2}{48 \sqrt{3} a_0^{3/2} N_2} + \frac{i a_1 a_2}{144 a_0^{3/2} N_2} \\
&\qquad + \frac{i a_3^2 N_1}{24 \sqrt{3} \pi a_0} - \frac{a_3^2 N_1}{72 \pi a_0} + \frac{i \sqrt{3} a_1 N_2}{4 \pi a_0^{5/6}} - \frac{3 a_1 N_2}{4 \pi a_0^{5/6}} - \frac{i a_3 a_4 N_2}{4 \sqrt{3} \pi a_0^{5/6}} + \frac{a_3 a_4 N_2}{4 \pi a_0^{5/6}} \\
&\qquad - \frac{a_1 a_3}{32 \sqrt{3} a_0^{4/3} N_1} - \frac{i a_1 a_3}{32 a_0^{4/3} N_1} + \frac{\sqrt{3} a_4}{32 \sqrt[3]{a_0} N_1} + \frac{3 i a_4}{32 \sqrt[3]{a_0} N_1} - \frac{a_3}{8 \sqrt{3} \sqrt{a_0} N_2} \\
&\qquad - \frac{i a_3}{24 \sqrt{a_0} N_2} + \frac{a_2}{6 a_0^{2/3}} - \frac{a_4^2}{24 a_0^{2/3}} + \frac{i \sqrt{3} N_1}{4 \pi} - \frac{N_1}{4 \pi} \Bigg) \\
&\quad + g \Bigg( -\frac{5 i a_1^2 N_1}{108 \sqrt{3} \pi a_0^{5/3}} + \frac{5 a_1^2 N_1}{324 \pi a_0^{5/3}} + \frac{7 i a_1 a_2 N_2}{72 \sqrt{3} \pi a_0^{3/2}} - \frac{7 a_1 a_2 N_2}{72 \pi a_0^{3/2}} + \frac{5 a_1 a_4}{432 \sqrt{3} a_0^{7/6} N_2} \\
&\qquad + \frac{5 i a_1 a_4}{1296 a_0^{7/6} N_2} + \frac{11 i a_4^2 N_1}{864 \sqrt{3} \pi a_0^{2/3}} - \frac{11 a_4^2 N_1}{2592 \pi a_0^{2/3}} - \frac{i a_2 N_1}{24 \sqrt{3} \pi a_0^{2/3}} + \frac{a_2 N_1}{72 \pi a_0^{2/3}} \\
&\qquad + \frac{i a_3 N_2}{24 \sqrt{3} \pi \sqrt{a_0}} - \frac{a_3 N_2}{24 \pi \sqrt{a_0}} - \frac{13 a_3^2}{1152 \sqrt{3} a_0 N_1} - \frac{13 i a_3^2}{1152 a_0 N_1} - \frac{11 a_2 a_4}{576 \sqrt{3} a_0 N_1} \\
&\qquad - \frac{11 i a_2 a_4}{576 a_0 N_1} + \frac{5 a_2 a_3}{288 \sqrt{3} a_0^{7/6} N_2} + \frac{5 i a_2 a_3}{864 a_0^{7/6} N_2} + \frac{5}{64 \sqrt{3} N_1} + \frac{5 i}{64 N_1} \Bigg) \\
&\quad + g^3 \Bigg( \frac{575 a_1^2}{5184 \sqrt{3} a_0^{5/3} N_1} + \frac{575 i a_1^2}{5184 a_0^{5/3} N_1} - \frac{707 i a_1 a_3 N_1}{3888 \sqrt{3} \pi a_0^{4/3}} + \frac{707 a_1 a_3 N_1}{11664 \pi a_0^{4/3}} \\
&\qquad + \frac{91 i a_1 a_4 N_2}{648 \sqrt{3} \pi a_0^{7/6}} - \frac{91 a_1 a_4 N_2}{648 \pi a_0^{7/6}} - \frac{275 a_1}{5184 \sqrt{3} a_0^{5/6} N_2} - \frac{275 i a_1}{15552 a_0^{5/6} N_2} \\
&\qquad - \frac{4907 i a_2^2 N_1}{38880 \sqrt{3} \pi a_0^{4/3}} + \frac{4907 a_2^2 N_1}{116640 \pi a_0^{4/3}} - \frac{41 i a_4 N_1}{1296 \sqrt{3} \pi \sqrt[3]{a_0}} + \frac{41 a_4 N_1}{3888 \pi \sqrt[3]{a_0}} \\
&\qquad + \frac{1183 i a_2 a_3 N_2}{4320 \sqrt{3} \pi a_0^{7/6}} - \frac{1183 a_2 a_3 N_2}{4320 \pi a_0^{7/6}} - \frac{1981 a_4^2}{82944 \sqrt{3} a_0^{2/3} N_1} - \frac{1981 i a_4^2}{82944 a_0^{2/3} N_1} \\
&\qquad + \frac{145 a_2}{2304 \sqrt{3} a_0^{2/3} N_1} + \frac{145 i a_2}{2304 a_0^{2/3} N_1} + \frac{649 a_3 a_4}{15552 \sqrt{3} a_0^{5/6} N_2} + \frac{649 i a_3 a_4}{46656 a_0^{5/6} N_2} \Bigg)
\end{align*}
\begin{align*}
B_2(g,a) &= \frac{1}{g} \Bigg( -\frac{a_2^2}{64 \sqrt{3} a_0^{4/3} N_1} + \frac{i a_2^2}{64 a_0^{4/3} N_1} - \frac{i a_2 a_4 N_1}{12 \sqrt{3} \pi a_0} - \frac{a_2 a_4 N_1}{36 \pi a_0} - \frac{a_1 a_2}{48 \sqrt{3} a_0^{3/2} N_2} + \frac{i a_1 a_2}{144 a_0^{3/2} N_2} \\
&\qquad - \frac{i a_3^2 N_1}{24 \sqrt{3} \pi a_0} - \frac{a_3^2 N_1}{72 \pi a_0} + \frac{i \sqrt{3} a_1 N_2}{4 \pi a_0^{5/6}} + \frac{3 a_1 N_2}{4 \pi a_0^{5/6}} - \frac{i a_3 a_4 N_2}{4 \sqrt{3} \pi a_0^{5/6}} - \frac{a_3 a_4 N_2}{4 \pi a_0^{5/6}} \\
&\qquad - \frac{a_1 a_3}{32 \sqrt{3} a_0^{4/3} N_1} + \frac{i a_1 a_3}{32 a_0^{4/3} N_1} + \frac{\sqrt{3} a_4}{32 \sqrt[3]{a_0} N_1} - \frac{3 i a_4}{32 \sqrt[3]{a_0} N_1} + \frac{a_3}{8 \sqrt{3} \sqrt{a_0} N_2} \\
&\qquad - \frac{i a_3}{24 \sqrt{a_0} N_2} + \frac{a_2}{6 a_0^{2/3}} - \frac{a_4^2}{24 a_0^{2/3}} - \frac{i \sqrt{3} N_1}{4 \pi} - \frac{N_1}{4 \pi} \Bigg) \\
&\quad + g \Bigg( \frac{5 i a_1^2 N_1}{108 \sqrt{3} \pi a_0^{5/3}} + \frac{5 a_1^2 N_1}{324 \pi a_0^{5/3}} + \frac{7 i a_1 a_2 N_2}{72 \sqrt{3} \pi a_0^{3/2}} + \frac{7 a_1 a_2 N_2}{72 \pi a_0^{3/2}} - \frac{5 a_1 a_4}{432 \sqrt{3} a_0^{7/6} N_2} \\
&\qquad + \frac{5 i a_1 a_4}{1296 a_0^{7/6} N_2} - \frac{11 i a_4^2 N_1}{864 \sqrt{3} \pi a_0^{2/3}} - \frac{11 a_4^2 N_1}{2592 \pi a_0^{2/3}} + \frac{i a_2 N_1}{24 \sqrt{3} \pi a_0^{2/3}} + \frac{a_2 N_1}{72 \pi a_0^{2/3}} \\
&\qquad + \frac{i a_3 N_2}{24 \sqrt{3} \pi \sqrt{a_0}} + \frac{a_3 N_2}{24 \pi \sqrt{a_0}} - \frac{13 a_3^2}{1152 \sqrt{3} a_0 N_1} + \frac{13 i a_3^2}{1152 a_0 N_1} - \frac{11 a_2 a_4}{576 \sqrt{3} a_0 N_1} \\
&\qquad + \frac{11 i a_2 a_4}{576 a_0 N_1} - \frac{5 a_2 a_3}{288 \sqrt{3} a_0^{7/6} N_2} + \frac{5 i a_2 a_3}{864 a_0^{7/6} N_2} + \frac{5}{64 \sqrt{3} N_1} - \frac{5 i}{64 N_1} \Bigg) \\
&\quad + g^3 \Bigg( \frac{575 a_1^2}{5184 \sqrt{3} a_0^{5/3} N_1} - \frac{575 i a_1^2}{5184 a_0^{5/3} N_1} + \frac{707 i a_1 a_3 N_1}{3888 \sqrt{3} \pi a_0^{4/3}} + \frac{707 a_1 a_3 N_1}{11664 \pi a_0^{4/3}} \\
&\qquad + \frac{91 i a_1 a_4 N_2}{648 \sqrt{3} \pi a_0^{7/6}} + \frac{91 a_1 a_4 N_2}{648 \pi a_0^{7/6}} + \frac{275 a_1}{5184 \sqrt{3} a_0^{5/6} N_2} - \frac{275 i a_1}{15552 a_0^{5/6} N_2} \\
&\qquad + \frac{4907 i a_2^2 N_1}{38880 \sqrt{3} \pi a_0^{4/3}} + \frac{4907 a_2^2 N_1}{116640 \pi a_0^{4/3}} + \frac{41 i a_4 N_1}{1296 \sqrt{3} \pi \sqrt[3]{a_0}} + \frac{41 a_4 N_1}{3888 \pi \sqrt[3]{a_0}} \\
&\qquad + \frac{1183 i a_2 a_3 N_2}{4320 \sqrt{3} \pi a_0^{7/6}} + \frac{1183 a_2 a_3 N_2}{4320 \pi a_0^{7/6}} - \frac{1981 a_4^2}{82944 \sqrt{3} a_0^{2/3} N_1} + \frac{1981 i a_4^2}{82944 a_0^{2/3} N_1} \\
&\qquad + \frac{145 a_2}{2304 \sqrt{3} a_0^{2/3} N_1} - \frac{145 i a_2}{2304 a_0^{2/3} N_1} - \frac{649 a_3 a_4}{15552 \sqrt{3} a_0^{5/6} N_2} + \frac{649 i a_3 a_4}{46656 a_0^{5/6} N_2} \Bigg)
\end{align*}
\begin{align*}
\mathcal{A}_1(g,a) &= \frac{1}{g} \Bigg( \frac{i \pi a_2^2}{32 \sqrt{3} a_0^{4/3} N_1} + \frac{\pi a_2^2}{32 a_0^{4/3} N_1} - \frac{a_2 a_4 N_1}{6 \sqrt{3} a_0} + \frac{i a_2 a_4 N_1}{18 a_0} - \frac{i \pi a_1 a_2}{24 \sqrt{3} a_0^{3/2} N_2} - \frac{\pi a_1 a_2}{72 a_0^{3/2} N_2} \\
&\qquad - \frac{a_3^2 N_1}{12 \sqrt{3} a_0} + \frac{i a_3^2 N_1}{36 a_0} - \frac{\sqrt{3} a_1 N_2}{2 a_0^{5/6}} + \frac{3 i a_1 N_2}{2 a_0^{5/6}} + \frac{a_3 a_4 N_2}{2 \sqrt{3} a_0^{5/6}} - \frac{i a_3 a_4 N_2}{2 a_0^{5/6}} \\
&\qquad + \frac{i \pi a_1 a_3}{16 \sqrt{3} a_0^{4/3} N_1} + \frac{\pi a_1 a_3}{16 a_0^{4/3} N_1} - \frac{i \sqrt{3} \pi a_4}{16 \sqrt[3]{a_0} N_1} - \frac{3 \pi a_4}{16 \sqrt[3]{a_0} N_1} + \frac{i \pi a_3}{4 \sqrt{3} \sqrt{a_0} N_2} \\
&\qquad + \frac{\pi a_3}{12 \sqrt{a_0} N_2} + \frac{2 i \pi a_2}{3 a_0^{2/3}} - \frac{i \pi a_4^2}{6 a_0^{2/3}} - \frac{\sqrt{3} N_1}{2} + \frac{i N_1}{2} \Bigg) \\
&\quad + g \Bigg( \frac{5 a_1^2 N_1}{54 \sqrt{3} a_0^{5/3}} - \frac{5 i a_1^2 N_1}{162 a_0^{5/3}} - \frac{7 a_1 a_2 N_2}{36 \sqrt{3} a_0^{3/2}} + \frac{7 i a_1 a_2 N_2}{36 a_0^{3/2}} - \frac{5 i \pi a_1 a_4}{216 \sqrt{3} a_0^{7/6} N_2} \\
&\qquad - \frac{5 \pi a_1 a_4}{648 a_0^{7/6} N_2} - \frac{11 a_4^2 N_1}{432 \sqrt{3} a_0^{2/3}} + \frac{11 i a_4^2 N_1}{1296 a_0^{2/3}} + \frac{a_2 N_1}{12 \sqrt{3} a_0^{2/3}} - \frac{i a_2 N_1}{36 a_0^{2/3}} \\
&\qquad - \frac{a_3 N_2}{12 \sqrt{3} \sqrt{a_0}} + \frac{i a_3 N_2}{12 \sqrt{a_0}} + \frac{13 i \pi a_3^2}{576 \sqrt{3} a_0 N_1} + \frac{13 \pi a_3^2}{576 a_0 N_1} + \frac{11 i \pi a_2 a_4}{288 \sqrt{3} a_0 N_1} \\
&\qquad + \frac{11 \pi a_2 a_4}{288 a_0 N_1} - \frac{5 i \pi a_2 a_3}{144 \sqrt{3} a_0^{7/6} N_2} - \frac{5 \pi a_2 a_3}{432 a_0^{7/6} N_2} - \frac{5 i \pi }{32 \sqrt{3} N_1} - \frac{5 \pi }{32 N_1} \Bigg) \\
&\quad + g^3 \Bigg( -\frac{575 i \pi a_1^2}{2592 \sqrt{3} a_0^{5/3} N_1} - \frac{575 \pi a_1^2}{2592 a_0^{5/3} N_1} + \frac{707 a_1 a_3 N_1}{1944 \sqrt{3} a_0^{4/3}} - \frac{707 i a_1 a_3 N_1}{5832 a_0^{4/3}} \\
&\qquad - \frac{91 a_1 a_4 N_2}{324 \sqrt{3} a_0^{7/6}} + \frac{91 i a_1 a_4 N_2}{324 a_0^{7/6}} + \frac{275 i \pi a_1}{2592 \sqrt{3} a_0^{5/6} N_2} + \frac{275 \pi a_1}{7776 a_0^{5/6} N_2} \\
&\qquad + \frac{4907 a_2^2 N_1}{19440 \sqrt{3} a_0^{4/3}} - \frac{4907 i a_2^2 N_1}{58320 a_0^{4/3}} + \frac{41 a_4 N_1}{648 \sqrt{3} \sqrt[3]{a_0}} - \frac{41 i a_4 N_1}{1944 \sqrt[3]{a_0}} \\
&\qquad - \frac{1183 a_2 a_3 N_2}{2160 \sqrt{3} a_0^{7/6}} + \frac{1183 i a_2 a_3 N_2}{2160 a_0^{7/6}} + \frac{1981 i \pi a_4^2}{41472 \sqrt{3} a_0^{2/3} N_1} + \frac{1981 \pi a_4^2}{41472 a_0^{2/3} N_1} \\
&\qquad - \frac{145 i \pi a_2}{1152 \sqrt{3} a_0^{2/3} N_1} - \frac{145 \pi a_2}{1152 a_0^{2/3} N_1} - \frac{649 i \pi a_3 a_4}{7776 \sqrt{3} a_0^{5/6} N_2} - \frac{649 \pi a_3 a_4}{23328 a_0^{5/6} N_2} \Bigg)
\end{align*}
\begin{align*}
\mathcal{A}_2(g,a) &= \frac{1}{g} \Bigg( -\frac{i \pi a_2^2}{32 \sqrt{3} a_0^{4/3} N_1} + \frac{\pi a_2^2}{32 a_0^{4/3} N_1} - \frac{a_2 a_4 N_1}{6 \sqrt{3} a_0} - \frac{i a_2 a_4 N_1}{18 a_0} - \frac{i \pi a_1 a_2}{24 \sqrt{3} a_0^{3/2} N_2} + \frac{\pi a_1 a_2}{72 a_0^{3/2} N_2} \\
&\qquad - \frac{a_3^2 N_1}{12 \sqrt{3} a_0} - \frac{i a_3^2 N_1}{36 a_0} + \frac{\sqrt{3} a_1 N_2}{2 a_0^{5/6}} + \frac{3 i a_1 N_2}{2 a_0^{5/6}} - \frac{a_3 a_4 N_2}{2 \sqrt{3} a_0^{5/6}} - \frac{i a_3 a_4 N_2}{2 a_0^{5/6}} \\
&\qquad - \frac{i \pi a_1 a_3}{16 \sqrt{3} a_0^{4/3} N_1} + \frac{\pi a_1 a_3}{16 a_0^{4/3} N_1} + \frac{i \sqrt{3} \pi a_4}{16 \sqrt[3]{a_0} N_1} - \frac{3 \pi a_4}{16 \sqrt[3]{a_0} N_1} + \frac{i \pi a_3}{4 \sqrt{3} \sqrt{a_0} N_2} \\
&\qquad - \frac{\pi a_3}{12 \sqrt{a_0} N_2} + \frac{i \pi a_2}{3 a_0^{2/3}} - \frac{i \pi a_4^2}{12 a_0^{2/3}} - \frac{\sqrt{3} N_1}{2} - \frac{i N_1}{2} \Bigg) \\
&\quad + g \Bigg( \frac{5 a_1^2 N_1}{54 \sqrt{3} a_0^{5/3}} + \frac{5 i a_1^2 N_1}{162 a_0^{5/3}} + \frac{7 a_1 a_2 N_2}{36 \sqrt{3} a_0^{3/2}} + \frac{7 i a_1 a_2 N_2}{36 a_0^{3/2}} - \frac{5 i \pi a_1 a_4}{216 \sqrt{3} a_0^{7/6} N_2} \\
&\qquad + \frac{5 \pi a_1 a_4}{648 a_0^{7/6} N_2} - \frac{11 a_4^2 N_1}{432 \sqrt{3} a_0^{2/3}} - \frac{11 i a_4^2 N_1}{1296 a_0^{2/3}} + \frac{a_2 N_1}{12 \sqrt{3} a_0^{2/3}} + \frac{i a_2 N_1}{36 a_0^{2/3}} \\
&\qquad + \frac{a_3 N_2}{12 \sqrt{3} \sqrt{a_0}} + \frac{i a_3 N_2}{12 \sqrt{a_0}} - \frac{13 i \pi a_3^2}{576 \sqrt{3} a_0 N_1} + \frac{13 \pi a_3^2}{576 a_0 N_1} - \frac{11 i \pi a_2 a_4}{288 \sqrt{3} a_0 N_1} \\
&\qquad + \frac{11 \pi a_2 a_4}{288 a_0 N_1} - \frac{5 i \pi a_2 a_3}{144 \sqrt{3} a_0^{7/6} N_2} + \frac{5 \pi a_2 a_3}{432 a_0^{7/6} N_2} + \frac{5 i \pi}{32 \sqrt{3} N_1} - \frac{5 \pi}{32 N_1} \Bigg) \\
&\quad + g^3 \Bigg( \frac{575 i \pi a_1^2}{2592 \sqrt{3} a_0^{5/3} N_1} - \frac{575 \pi a_1^2}{2592 a_0^{5/3} N_1} + \frac{707 a_1 a_3 N_1}{1944 \sqrt{3} a_0^{4/3}} + \frac{707 i a_1 a_3 N_1}{5832 a_0^{4/3}} \\
&\qquad + \frac{91 a_1 a_4 N_2}{324 \sqrt{3} a_0^{7/6}} + \frac{91 i a_1 a_4 N_2}{324 a_0^{7/6}} + \frac{275 i \pi a_1}{2592 \sqrt{3} a_0^{5/6} N_2} - \frac{275 \pi a_1}{7776 a_0^{5/6} N_2} \\
&\qquad + \frac{4907 a_2^2 N_1}{19440 \sqrt{3} a_0^{4/3}} + \frac{4907 i a_2^2 N_1}{58320 a_0^{4/3}} + \frac{41 a_4 N_1}{648 \sqrt{3} \sqrt[3]{a_0}} + \frac{41 i a_4 N_1}{1944 \sqrt[3]{a_0}} \\
&\qquad + \frac{1183 a_2 a_3 N_2}{2160 \sqrt{3} a_0^{7/6}} + \frac{1183 i a_2 a_3 N_2}{2160 a_0^{7/6}} - \frac{1981 i \pi a_4^2}{41472 \sqrt{3} a_0^{2/3} N_1} + \frac{1981 \pi a_4^2}{41472 a_0^{2/3} N_1} \\
&\qquad + \frac{145 i \pi a_2}{1152 \sqrt{3} a_0^{2/3} N_1} - \frac{145 \pi a_2}{1152 a_0^{2/3} N_1} - \frac{649 i \pi a_3 a_4}{7776 \sqrt{3} a_0^{5/6} N_2} + \frac{649 \pi a_3 a_4}{23328 a_0^{5/6} N_2} \Bigg)
\end{align*}

\endgroup

Here, $N_1=B(1/6,3/2)/3$ and $N_2=-\Gamma(1/3) \Gamma(3/2)/(3 \Gamma(5/6))$.
We ca also construct the periods by choosing another cycles such as $\mathfrak{a}_1=-\gamma_0-\gamma_1-\gamma_2$, $\mathfrak{a}_2=-\gamma_2+\gamma_3+\gamma_4$, $\mathfrak{b}_1=\gamma_0$, $\mathfrak{b}_2=\gamma_4$ ,or $\mathfrak{a}_1=-\gamma_2$, $\mathfrak{a}_2=-\gamma_1-\gamma_3$, $\mathfrak{b}_1=-\gamma_2-\gamma_3$, $\mathfrak{b}_2=\gamma_0-\gamma_1-\gamma_3$.

Let us use the shorthand notation for the partial derivatives with respect to moduli parameters $\partial_{n_0n_1n_2n_3n_4}=\partial_{u}^{n_0}\partial_{a_1}^{n_1}\partial_{a_2}^{n_2}\partial_{a_3}^{n_3}\partial_{a_4}^{n_4}$ for simplicity. We use the following quantum correction,
\begin{dmath}
\mathcal{D}_2 = \frac{5}{2}\partial_{10001}+a_4\partial_{10100}+\frac{a_3}{2}\partial_{11000}-\frac{a_2}{6}\partial_{20000}
\end{dmath}
\begin{dmath}
\mathcal{D}_4=\frac{35}{8}\partial_{20002}+\frac{5}{4}\partial_{20100}+\frac{7a_4}{2}\partial_{20101}+\frac{7a_3}{4}\partial_{21001}-\frac{a_4}{12}\partial_{30000}-\bigg(\frac{7a_2}{12}+\frac{7a_4^2}{10}\bigg)\partial_{30001}-\frac{7a_3a_4}{10}\partial_{30010}-\bigg(\frac{7a_2a_4}{30}+\frac{7a_3^2}{40}\bigg)\partial_{30100}-\frac{7a_2a_3}{60}\partial_{31000}+\frac{7a_2^2}{360}\partial_{40000}
\end{dmath}
\begin{dmath}
\mathcal{D}_6=\frac{775}{112}\partial_{30003}+\frac{2995}{504}\partial_{30101}+\frac{465a_4}{56}\partial_{30102}+\frac{465a_3}{112}\partial_{31002}-\frac{5}{72}\partial_{40000}-\frac{1387a_4}{504}\partial_{40001}-\bigg(\frac{155a_2}{112}+\frac{93a_4^2}{28}\bigg)\partial_{40002}-\frac{583a_3}{504}\partial_{40010}-\frac{93a_3a_4}{28}\partial_{40011}-\bigg(\frac{3a_2}{8}+\frac{41a_4^2}{252}\bigg)\partial_{40100}-\bigg(\frac{31a_2a_4}{28}+\frac{93a_3^2}{112}+\frac{31a_4^3}{70}\bigg)\partial_{40101}-\frac{41a_3a_4}{504}\partial_{41000}-\frac{31a_3(5a_2+6a_4^2)}{280}\partial_{41001}+\bigg(\frac{a_2a_4}{40}+\frac{a_3^2}{1260}\bigg)\partial_{50000}+\bigg(\frac{31a_2^2}{336}+\frac{31a_2a_4^2}{140}+\frac{93a_3^2a_4}{280}\bigg)\partial_{50001}+\frac{31a_3(4a_2a_4+3a_3^2)}{560}\partial_{50010}+\frac{31a_2(2a_2a_4+3a_3^2)}{1680}\partial_{50100}+\frac{31a_2^2a_3}{1680}\partial_{51000}-\frac{31a_2^3}{15120}\partial_{60000}
\end{dmath}

\subsection{Rational potential}
We give the Picard-Fuchs equations for the rational genus-2 curve of interest and then solve the Picard-Fuchs equations and obtain the periods, up to total moduli parameter order 4. We write a linear combination of periods such that their cycles become $\mathfrak{a}$ and $\mathfrak{b}$ cycles. 
\begin{equation}
\begin{split}
\partial_{\xi}\partial_{\alpha}\Pi+\partial_{\beta}^2\Pi+\gamma^2\partial_{\alpha}^2\Pi=0,\\
2\partial^2_{\xi}\Pi +2(\xi+\gamma^2)\partial_{\xi}\partial_{\alpha}\Pi+2\alpha\partial^2_{\alpha}\Pi+2\beta\partial_{\alpha}\partial_{\beta}\Pi+\partial_{\alpha}\Pi =0,\\
2(\xi+\gamma^2)\partial_{\xi}\partial_{\beta}\Pi+4\alpha \partial_{\alpha}\partial_{\beta}\Pi+3\beta \partial^2_{\beta}\Pi-\gamma^2\beta\partial^2_{\alpha}\Pi+\partial_{\beta}\Pi=0,\\
2\xi\partial^2_{\xi}\Pi+4\alpha\partial_{\xi}\partial_{\alpha}\Pi+3\beta\partial_{\xi}\partial_{\beta}\Pi+2\gamma^2\alpha\partial^2_{\alpha}\Pi+2\gamma^2\beta\partial_{\alpha}\partial_{\beta}\Pi=0,\\
4\beta(\xi-\gamma^2)\partial_{\xi}\partial_{\beta}\Pi+(4\alpha(\xi-\gamma^2)-6\beta^2)\partial_{\xi}\partial_{\alpha}\Pi
+2\beta(7\alpha-2\gamma^2(\xi+\gamma^2))\partial_{\alpha}\partial_{\beta}\Pi \qquad\\
+(8\alpha^2-4\alpha\gamma^2(\xi+\gamma^2)-6\beta^2\gamma^2)\partial^2_{\alpha}\Pi-2\xi\partial_{\xi}\Pi-\beta\partial_{\beta}\Pi+2\Pi=0
\end{split}
\end{equation}

Also, we have the weight operator, given as follows
\begin{equation}
    4\alpha \partial_{\alpha}\Pi+3\beta\partial_{\beta}\Pi+\gamma\partial_{\gamma}\Pi+2\xi\partial_{\xi}\Pi-2\Pi=0,
\end{equation}
which implis that the terms is expansion of the periods should be of the form below.
\begin{equation}
    \gamma^2 C_{ijk} \left ( \frac{\alpha}{\gamma^4} \right )^i \left ( \frac{\beta}{\gamma^3} \right )^j \left ( \frac{\xi}{\gamma^2} \right )^k
\end{equation}
We used the weight-free parameter notation $ a=\frac{\alpha}{\gamma^4}$, $b=\frac{\beta}{\gamma^3}$, $x=\frac{\xi}{\gamma^2}$, $h=\frac{g}{\gamma^2}$ for simplicity.
\begingroup\makeatletter\def\f@size{10}\check@mathfonts

\begin{align*}
&B_1 = \frac{\gamma^2 x}{6}+ \frac{\gamma^2\Gamma\!\left(\frac13\right)^3}{2^{4/3}\pi^2}\Biggl[
 -\tfrac{7}{240} - \tfrac{29}{54}\,a - \tfrac{5561}{13824}\,b + \tfrac{3}{8}\,x + \tfrac{47}{216}\,a^{2} + \tfrac{59}{384}\,ab - \tfrac{13}{216}\,b^{2} - \tfrac{17}{54}\,ax - \tfrac{179}{3456}\,bx + \tfrac{7}{48}\,x^{2} 
 \\[-1pt] &- \tfrac{1}{9}\,a^{3} - \tfrac{245}{2304}\,a^{2}b + \tfrac{1}{48}\,ab^{2} + \tfrac{1}{576}\,b^{3} + \tfrac{5}{16}\,a^{2}x + \tfrac{233}{1152}\,abx + \tfrac{5}{144}\,b^{2}x - \tfrac{25}{72}\,ax^{2} - \tfrac{341}{2304}\,bx^{2} + \tfrac{7}{48}\,x^{3}
\\[-1pt] & +h^2\Bigl(
 \tfrac{7}{48} - \tfrac{13}{108}\,a - \tfrac{23}{1728}\,b + \tfrac{7}{144}\,x + \tfrac{53}{432}\,a^{2} + \tfrac{35}{288}\,ab + \tfrac{23}{1728}\,b^{2} - \tfrac{269}{864}\,ax - \tfrac{55}{288}\,bx + \tfrac{35}{144}\,x^{2} + \tfrac{29}{1296}\,a^{3}
 \\[-1pt] &+ \tfrac{7}{64}\,a^{2}b + \tfrac{1}{8}\,ab^{2} + \tfrac{385}{10368}\,b^{3} - \tfrac{71}{432}\,a^{2}x - \tfrac{119}{288}\,abx - \tfrac{23}{108}\,b^{2}x + \tfrac{49}{144}\,ax^{2} + \tfrac{665}{1728}\,bx^{2} - \tfrac{259}{1152}\,x^{3}
\Bigr)\Biggr]
\\[2pt] &+ \frac{2^{4/3}\sqrt3\,\pi\gamma^2}{\Gamma\!\left(\frac13\right)^3}\Biggl[
 -\tfrac{1}{9} - \tfrac{13}{81}\,a - \tfrac{11}{2592}\,b + \tfrac{13}{96}\,x + \tfrac{17}{162}\,a^{2} - \tfrac{2}{27}\,ab - \tfrac{53}{1296}\,b^{2}- \tfrac{83}{648}\,ax + \tfrac{461}{5184}\,bx + \tfrac{5}{144}\,x^{2} 
 \\[-1pt] &- \tfrac{1}{36}\,a^{3} + \tfrac{5}{54}\,a^{2}b + \tfrac{11}{144}\,ab^{2} + \tfrac{25}{3456}\,b^{3} + \tfrac{1}{144}\,a^{2}x - \tfrac{395}{1728}\,abx - \tfrac{37}{432}\,b^{2}x + \tfrac{17}{216}\,ax^{2} + \tfrac{271}{1728}\,bx^{2} - \tfrac{5}{72}\,x^{3}
\\[-1pt] & +h^2\Bigl(
 \tfrac{5}{432} - \tfrac{31}{162}\,a - \tfrac{133}{1296}\,b + \tfrac{125}{432}\,x + \tfrac{73}{1296}\,a^{2} - \tfrac{5}{72}\,ab - \tfrac{31}{648}\,b^{2} - \tfrac{5}{216}\,ax + \tfrac{245}{1296}\,bx - \tfrac{85}{864}\,x^{2} - \tfrac{65}{972}\,a^{3}
 \\[-1pt] &- \tfrac{55}{432}\,a^{2}b - \tfrac{25}{216}\,ab^{2} - \tfrac{295}{7776}\,b^{3} + \tfrac{65}{216}\,a^{2}x + \tfrac{65}{144}\,abx + \tfrac{185}{864}\,b^{2}x - \tfrac{1265}{2592}\,ax^{2} - \tfrac{5}{12}\,bx^{2} + \tfrac{245}{864}\,x^{3}
\Bigr)\Biggr]+\mathcal{O}(4)
\end{align*}
\begin{align*}
&B_2 = \frac{\gamma^2 x}{6}
+ \frac{\gamma^2\Gamma\!\left(\frac13\right)^3}{2^{4/3}\pi^2}\Biggl[
 -\tfrac{7}{240} - \tfrac{29}{54}\,a + \tfrac{5561}{13824}\,b + \tfrac{3}{8}\,x + \tfrac{47}{216}\,a^{2} - \tfrac{59}{384}\,ab - \tfrac{13}{216}\,b^{2}  - \tfrac{17}{54}\,ax + \tfrac{179}{3456}\,bx + \tfrac{7}{48}\,x^{2} 
 \\[-1pt] &- \tfrac{1}{9}\,a^{3} + \tfrac{245}{2304}\,a^{2}b + \tfrac{1}{48}\,ab^{2} - \tfrac{1}{576}\,b^{3} + \tfrac{5}{16}\,a^{2}x - \tfrac{233}{1152}\,abx + \tfrac{5}{144}\,b^{2}x - \tfrac{25}{72}\,ax^{2} + \tfrac{341}{2304}\,bx^{2}  + \tfrac{7}{48}\,x^{3}
\\[-1pt] & +h^2\Bigl(
 \tfrac{7}{48} - \tfrac{13}{108}\,a + \tfrac{23}{1728}\,b + \tfrac{7}{144}\,x + \tfrac{53}{432}\,a^{2} - \tfrac{35}{288}\,ab + \tfrac{23}{1728}\,b^{2} - \tfrac{269}{864}\,ax + \tfrac{55}{288}\,bx + \tfrac{35}{144}\,x^{2} + \tfrac{29}{1296}\,a^{3} 
 \\[-1pt] & - \tfrac{7}{64}\,a^{2}b + \tfrac{1}{8}\,ab^{2}- \tfrac{385}{10368}\,b^{3} - \tfrac{71}{432}\,a^{2}x + \tfrac{119}{288}\,abx - \tfrac{23}{108}\,b^{2}x + \tfrac{49}{144}\,ax^{2}  - \tfrac{665}{1728}\,bx^{2} - \tfrac{259}{1152}\,x^{3}
\Bigr)\Biggr]
\\[2pt] &+ \frac{2^{4/3}\sqrt3\,\pi\gamma^2}{\Gamma\!\left(\frac13\right)^3}\Biggl[
 -\tfrac{1}{9} - \tfrac{13}{81}\,a + \tfrac{11}{2592}\,b + \tfrac{13}{96}\,x + \tfrac{17}{162}\,a^{2} + \tfrac{2}{27}\,ab - \tfrac{53}{1296}\,b^{2}  - \tfrac{83}{648}\,ax - \tfrac{461}{5184}\,bx + \tfrac{5}{144}\,x^{2}\\[-1pt] & - \tfrac{1}{36}\,a^{3} - \tfrac{5}{54}\,a^{2}b + \tfrac{11}{144}\,ab^{2} - \tfrac{25}{3456}\,b^{3} + \tfrac{1}{144}\,a^{2}x + \tfrac{395}{1728}\,abx - \tfrac{37}{432}\,b^{2}x + \tfrac{17}{216}\,ax^{2} - \tfrac{271}{1728}\,bx^{2} - \tfrac{5}{72}\,x^{3}
\\[-1pt] & +h^2\Bigl(
 \tfrac{5}{432} - \tfrac{31}{162}\,a + \tfrac{133}{1296}\,b + \tfrac{125}{432}\,x + \tfrac{73}{1296}\,a^{2} + \tfrac{5}{72}\,ab - \tfrac{31}{648}\,b^{2} - \tfrac{5}{216}\,ax - \tfrac{245}{1296}\,bx - \tfrac{85}{864}\,x^{2} - \tfrac{65}{972}\,a^{3}
 \\[-1pt] &+ \tfrac{55}{432}\,a^{2}b - \tfrac{25}{216}\,ab^{2}+ \tfrac{295}{7776}\,b^{3} + \tfrac{65}{216}\,a^{2}x - \tfrac{65}{144}\,abx + \tfrac{185}{864}\,b^{2}x - \tfrac{1265}{2592}\,ax^{2} + \tfrac{5}{12}\,bx^{2}
 + \tfrac{245}{864}\,x^{3}
\Bigr)\Biggr]+\mathcal{O}(4)
\end{align*}
\begin{align*}
&\mathcal{A}_1 = -\frac{\gamma^2 x}{6}+ \frac12\bigl(1+i\sqrt3\bigr)\frac{\gamma^2\Gamma\!\left(\frac13\right)^3}{2^{4/3}\pi^2}\Biggl[
 -\tfrac{7}{240} - \tfrac{29}{54}\,a + \tfrac{3}{8}\,x + \tfrac{47}{216}\,a^{2} - \tfrac{13}{216}\,b^{2} - \tfrac{17}{54}\,ax + \tfrac{7}{48}\,x^{2} - \tfrac{1}{9}\,a^{3}
\\[-1pt] &  + \tfrac{1}{48}\,ab^{2} + \tfrac{5}{16}\,a^{2}x + \tfrac{5}{144}\,b^{2}x - \tfrac{25}{72}\,ax^{2} + \tfrac{7}{48}\,x^{3}+h^2\Bigl(
 \tfrac{7}{48} - \tfrac{13}{108}\,a + \tfrac{7}{144}\,x + \tfrac{53}{432}\,a^{2} + \tfrac{23}{1728}\,b^{2} - \tfrac{269}{864}\,ax 
 \\[-1pt] &+ \tfrac{35}{144}\,x^{2} + \tfrac{29}{1296}\,a^{3} + \tfrac{1}{8}\,ab^{2} - \tfrac{71}{432}\,a^{2}x  - \tfrac{23}{108}\,b^{2}x + \tfrac{49}{144}\,ax^{2} - \tfrac{259}{1152}\,x^{3}
\Bigr)\Biggr]
\\[2pt] &+ \frac12\bigl(-1+i\sqrt3\bigr)\frac{\gamma^2\Gamma\!\left(\frac13\right)^3}{2^{4/3}\pi^2}\Biggl[
 -\tfrac{5561}{13824}\,b + \tfrac{59}{384}\,ab - \tfrac{179}{3456}\,bx - \tfrac{245}{2304}\,a^{2}b + \tfrac{1}{576}\,b^{3} + \tfrac{233}{1152}\,abx - \tfrac{341}{2304}\,bx^{2}
\\[-1pt] & {}+h^2\Bigl(
 -\tfrac{23}{1728}\,b + \tfrac{35}{288}\,ab - \tfrac{55}{288}\,bx + \tfrac{7}{64}\,a^{2}b + \tfrac{385}{10368}\,b^{3} - \tfrac{119}{288}\,abx + \tfrac{665}{1728}\,bx^{2}
\Bigr)\Biggr]
\\[2pt] &+ \frac12\bigl(1-i\sqrt3\bigr)\frac{2^{4/3}\sqrt3\,\pi\gamma^2}{\Gamma\!\left(\frac13\right)^3}\Biggl[
 -\tfrac{1}{9} - \tfrac{13}{81}\,a + \tfrac{13}{96}\,x + \tfrac{17}{162}\,a^{2} - \tfrac{53}{1296}\,b^{2} - \tfrac{83}{648}\,ax + \tfrac{5}{144}\,x^{2} - \tfrac{1}{36}\,a^{3} + \tfrac{11}{144}\,ab^{2} 
 \\[-1pt] & + \tfrac{1}{144}\,a^{2}x - \tfrac{37}{432}\,b^{2}x + \tfrac{17}{216}\,ax^{2} - \tfrac{5}{72}\,x^{3}+h^2\Bigl(
 \tfrac{5}{432} - \tfrac{31}{162}\,a + \tfrac{125}{432}\,x + \tfrac{73}{1296}\,a^{2} - \tfrac{31}{648}\,b^{2} - \tfrac{5}{216}\,ax - \tfrac{85}{864}\,x^{2}
\\[-1pt] &  - \tfrac{65}{972}\,a^{3} - \tfrac{25}{216}\,ab^{2} + \tfrac{65}{216}\,a^{2}x + \tfrac{185}{864}\,b^{2}x - \tfrac{1265}{2592}\,ax^{2}+ \tfrac{245}{864}\,x^{3}
\Bigr)\Biggr]
\\[2pt] &+ \frac12\bigl(-1-i\sqrt3\bigr)\frac{2^{4/3}\sqrt3\,\pi\gamma^2}{\Gamma\!\left(\frac13\right)^3}\Biggl[
 -\tfrac{11}{2592}\,b - \tfrac{2}{27}\,ab + \tfrac{461}{5184}\,bx + \tfrac{5}{54}\,a^{2}b + \tfrac{25}{3456}\,b^{3} - \tfrac{395}{1728}\,abx + \tfrac{271}{1728}\,bx^{2}
\\[-1pt] & +h^2\Bigl(
 -\tfrac{133}{1296}\,b - \tfrac{5}{72}\,ab + \tfrac{245}{1296}\,bx - \tfrac{55}{432}\,a^{2}b - \tfrac{295}{7776}\,b^{3} + \tfrac{65}{144}\,abx  - \tfrac{5}{12}\,bx^{2}
\Bigr)\Biggr]+\mathcal{O}(4)
\end{align*}
\begin{align*}
&\mathcal{A}_2 = -\frac{\gamma^2 x}{6}+ \frac12\bigl(1+i\sqrt3\bigr)\frac{\gamma^2\Gamma\!\left(\frac13\right)^3}{2^{4/3}\pi^2}\Biggl[
 -\tfrac{7}{240} - \tfrac{29}{54}\,a + \tfrac{3}{8}\,x + \tfrac{47}{216}\,a^{2} - \tfrac{13}{216}\,b^{2} - \tfrac{17}{54}\,ax + \tfrac{7}{48}\,x^{2} - \tfrac{1}{9}\,a^{3}
\\[-1pt] &  + \tfrac{1}{48}\,ab^{2} + \tfrac{5}{16}\,a^{2}x + \tfrac{5}{144}\,b^{2}x - \tfrac{25}{72}\,ax^{2} + \tfrac{7}{48}\,x^{3}+h^2\Bigl(
 \tfrac{7}{48} - \tfrac{13}{108}\,a + \tfrac{7}{144}\,x + \tfrac{53}{432}\,a^{2} + \tfrac{23}{1728}\,b^{2}
 \\[-1pt] & - \tfrac{269}{864}\,ax + \tfrac{35}{144}\,x^{2}   + \tfrac{29}{1296}\,a^{3} + \tfrac{1}{8}\,ab^{2} - \tfrac{71}{432}\,a^{2}x - \tfrac{23}{108}\,b^{2}x + \tfrac{49}{144}\,ax^{2} - \tfrac{259}{1152}\,x^{3}
\Bigr)\Biggr]
\\[2pt] &+ \frac12\bigl(1-i\sqrt3\bigr)\frac{\gamma^2\Gamma\!\left(\frac13\right)^3}{2^{4/3}\pi^2}\Biggl[
 -\tfrac{5561}{13824}\,b + \tfrac{59}{384}\,ab - \tfrac{179}{3456}\,bx - \tfrac{245}{2304}\,a^{2}b + \tfrac{1}{576}\,b^{3} + \tfrac{233}{1152}\,abx - \tfrac{341}{2304}\,bx^{2}
 \\[-1pt] & +h^2\Bigl(
 -\tfrac{23}{1728}\,b + \tfrac{35}{288}\,ab - \tfrac{55}{288}\,bx + \tfrac{7}{64}\,a^{2}b + \tfrac{385}{10368}\,b^{3} - \tfrac{119}{288}\,abx  + \tfrac{665}{1728}\,bx^{2}
\Bigr)\Biggr]
\\[2pt] &+ \frac12\bigl(1-i\sqrt3\bigr)\frac{2^{4/3}\sqrt3\,\pi\gamma^2}{\Gamma\!\left(\frac13\right)^3}\Biggl[
 -\tfrac{1}{9} - \tfrac{13}{81}\,a + \tfrac{13}{96}\,x + \tfrac{17}{162}\,a^{2} - \tfrac{53}{1296}\,b^{2} - \tfrac{83}{648}\,ax + \tfrac{5}{144}\,x^{2}
\\[-1pt] &  - \tfrac{1}{36}\,a^{3} + \tfrac{11}{144}\,ab^{2} + \tfrac{1}{144}\,a^{2}x - \tfrac{37}{432}\,b^{2}x + \tfrac{17}{216}\,ax^{2} - \tfrac{5}{72}\,x^{3}+h^2\Bigl(
 \tfrac{5}{432} - \tfrac{31}{162}\,a + \tfrac{125}{432}\,x + \tfrac{73}{1296}\,a^{2} 
 \\[-1pt] & - \tfrac{31}{648}\,b^{2} - \tfrac{5}{216}\,ax - \tfrac{85}{864}\,x^{2} - \tfrac{65}{972}\,a^{3} - \tfrac{25}{216}\,ab^{2} + \tfrac{65}{216}\,a^{2}x + \tfrac{185}{864}\,b^{2}x - \tfrac{1265}{2592}\,ax^{2} + \tfrac{245}{864}\,x^{3}
\Bigr)\Biggr]
\\[2pt] &+ \frac12\bigl(1+i\sqrt3\bigr)\frac{2^{4/3}\sqrt3\,\pi\gamma^2}{\Gamma\!\left(\frac13\right)^3}\Biggl[
 -\tfrac{11}{2592}\,b - \tfrac{2}{27}\,ab + \tfrac{461}{5184}\,bx + \tfrac{5}{54}\,a^{2}b + \tfrac{25}{3456}\,b^{3} - \tfrac{395}{1728}\,abx
\\[-1pt] &  + \tfrac{271}{1728}\,bx^{2}+h^2\Bigl(
 -\tfrac{133}{1296}\,b - \tfrac{5}{72}\,ab + \tfrac{245}{1296}\,bx - \tfrac{55}{432}\,a^{2}b - \tfrac{295}{7776}\,b^{3} + \tfrac{65}{144}\,abx  - \tfrac{5}{12}\,bx^{2}
\Bigr)\Biggr]+\mathcal{O}(4)
\end{align*}
\endgroup

 \subsection{Non-hyperelliptic genus-3}

We checked the P/NP relation for non-hyperelliptic genus-3 example up to total $u^{n_1}\beta^{n_2}\gamma^{n_3}$ order $n_1+n_2+n_3\leq3$, and up the first quantum-correction.

By differentiating the curve we have, we obtained the following Picard-Fuchs equations, for the non-scaled (raw) periods of the curve.

\begin{equation}\label{PFNHG3}
\begin{split}
    \partial_u^2 \Pi+\frac u3\partial_{\gamma}^2 \Pi =0,\\
    \partial_{\alpha}\partial_{\gamma}\Pi -\partial_{\beta}^2\Pi=0,\\
    4 \partial_{\alpha} \partial_{\beta}\Pi+2\alpha \partial_{\beta}\partial_{\gamma}\Pi+\beta\partial_{\gamma}^2\Pi=0,\\
    4\partial_{\alpha}^2\Pi+2\alpha \partial_{\alpha}\partial_{\gamma}\Pi+\beta\partial_{\beta}\partial_{\gamma}\Pi+\partial_\gamma \Pi=0,\\
    8u\partial_u \Pi + 6\alpha \partial_{\alpha}\Pi+9\beta \partial_{\beta}+12 \gamma\partial_{\gamma}\Pi-7\Pi=0.
    \end{split}
\end{equation}
It is obvious that the last equation in \eqref{PFNHG3} is nothing but weight operator, which implies that the terms in our expansion should have the following form,
\begin{equation}\label{NHG3weighted}
    \alpha^{7/6}C_{ijk}\left ( \frac{u}{\alpha^{4/3}}\right )^i \left ( \frac{\beta}{\alpha^{3/2}}\right )^j \left ( \frac{\gamma}{\alpha^{2}}\right )^k.
\end{equation}
Then, we write a power expansion which is a sum of the terms of the form above, and solved the coefficients by using the PF equations. We also checked that our period coincides with the periods of the curve $y^3+z^4+\alpha z^2+\beta z+ \gamma=0$, which is the $v\rightarrow0$ limit of our original curve. Then we take the linear combination of the periods that solve the Picard-Fuchs equations \eqref{PFNHG3}, such that the corresponding cycles obey our $\mathfrak{a}$ and $\mathfrak{b}$ conventions. Hence, we obtain the perturbative and non-perturbative periods as power expansions of the moduli parameters. We used the weight-free parameter notation $b=\frac{\beta}{\alpha^{3/2}}$, $c=\frac{\gamma}{\alpha^2}$, $u=\frac{v}{\alpha^{4/3}}$, $h=\frac{g}{\alpha^{7/6}}$, and denoted the complex coefficients as $\omega=\frac{-1+i\sqrt3}{2}$, $\rho=2^{1/3}(\sqrt3-i)$ for simplicity.
\begingroup\makeatletter\def\f@size{10}\check@mathfonts
\begin{align*}
&B_1 ={}\frac{6\pi\bigl(i+\frac{1}{\sqrt3}\bigr)}{7\,2^{1/3}\Gamma\!\bigl(\frac13\bigr)^3}\Biggl\{\frac{1}{h}\Bigl[
1 + \tfrac{7}{6}c + \tfrac{55}{72}c^{2} + \tfrac{21505}{16848}c^{3} + \tfrac{21827575}{7682688}c^{4} + \tfrac{1682469481}{230480640}c^{5} - b^{2}\bigl(\tfrac{35}{72} + \tfrac{935}{432}c + \tfrac{623645}{67392}c^{2}\\[-1pt] & + \tfrac{894930575}{23048064}c^{3}\bigr) \; + b^{4}\bigl(\tfrac{21505}{31104} + \tfrac{21827575}{2426112}c\bigr) - \rho b u\bigl(\tfrac{7}{36} + \tfrac{35}{54}c + \tfrac{385}{162}c^{2} + \tfrac{6545}{729}c^{3}\bigr) \; + \rho b^{3} u\bigl(\tfrac{385}{972} + \tfrac{6545}{1458}c\bigr)
\\[-1pt] & - u^{3}\bigl(\tfrac{55}{648} + \tfrac{21505}{50544}c + \tfrac{21827575}{11524032}c^{2}\bigr) \; + b^{2} u^{3}\bigl(\tfrac{623645}{606528}\bigr) \; + \rho b u^{4}\bigl(\tfrac{385}{2916}\bigr)
\Bigr] +h\Bigl[
\rho b\bigl(\tfrac{35}{54} + \tfrac{385}{81}c + \tfrac{6545}{243}c^{2} + \tfrac{150535}{729}c^{3}\bigr)
\\[-1pt] & - \rho b^{3}\bigl(\tfrac{9625}{2916} + \tfrac{65450}{729}c\bigr) \; + u^{2}\bigl(\tfrac{51425}{33696} + \tfrac{49267955}{3841344}c + \tfrac{894930575}{7682688}c^{2}\bigr) \; - b^{2} u^{2}\bigl(\tfrac{21827575}{404352}\bigr)  - \rho b u^{3}\bigl(\tfrac{6545}{729}\bigr)
\Bigr]\Biggr\}
\\
&+ \frac{(\sqrt3-i)\Gamma\!\bigl(\frac13\bigr)^3}{16\pi^2}\Biggl\{\frac{1}{h}\Bigl[
b\bigl(1 + \tfrac{2}{3}c + \tfrac{14}{9}c^{2} + \tfrac{364}{81}c^{3} + \tfrac{3458}{243}c^{4}\bigr) - b^{3}\bigl(\tfrac{7}{27} + \tfrac{182}{81}c + \tfrac{3458}{243}c^{2}\bigr) \; + b^{5}\bigl(\tfrac{728}{1215}\bigr)
\\[-1pt] & + \rho u\bigl(\tfrac{1}{2} + \tfrac{7}{60}c + \tfrac{1729}{7920}c^{2} + \tfrac{267995}{484704}c^{3} + \tfrac{426380045}{267556608}c^{4}\bigr) \; - \rho b^{2} u\bigl(\tfrac{91}{720} + \tfrac{8645}{9504}c + \tfrac{9915815}{1938816}c^{2}\bigr)
\\[-1pt] & + \rho b^{4} u\bigl(\tfrac{267995}{684288}\bigr) \; - b u^{3}\bigl(\tfrac{14}{81} + \tfrac{364}{243}c\bigr) \; - \rho u^{4}\bigl(\tfrac{1729}{142560} + \tfrac{267995}{2908224}c\bigr)
\Bigr]+h\Bigl[
-\rho\bigl(\tfrac{23}{120} + \tfrac{4277}{7920}c + \tfrac{613795}{323136}c^{2} 
\\[-1pt] &+ \tfrac{942002425}{133778304}c^{3} + \tfrac{20892622205}{535113216}c^{4}\bigr)
 + \rho b^{2}\bigl(\tfrac{70889}{95040} + \tfrac{17419675}{1938816}c + \tfrac{426380045}{3877632}c^{2}\bigr) \; - \rho b^{4}\bigl(\tfrac{9915815}{1368576}\bigr) \; + b u^{2}\bigl(\tfrac{364}{81} + \tfrac{6916}{81}c\bigr)
\\[-1pt] & + \rho u^{3}\bigl(\tfrac{613795}{1454112} + \tfrac{9915815}{1454112}c\bigr)
\Bigr]\Biggr\}+\mathcal{O}(6)
\end{align*}

\begin{align*}
&B_2 =\frac{6\pi\bigl(i+\frac{1}{\sqrt3}\bigr)}{7\,2^{1/3}\Gamma\!\bigl(\frac13\bigr)^3}\Biggl\{\frac{1}{h}\Bigl[
\omega\bigl(1 + \tfrac{7}{6}c + \tfrac{55}{72}c^{2} + \tfrac{21505}{16848}c^{3} + \tfrac{21827575}{7682688}c^{4} + \tfrac{1682469481}{230480640}c^{5}\bigr) - \omega b^{2}\bigl(\tfrac{35}{72} + \tfrac{935}{432}c + \tfrac{623645}{67392}c^{2}\\[-1pt] &
+ \tfrac{894930575}{23048064}c^{3}\bigr) \; + \omega b^{4}\bigl(\tfrac{21505}{31104} + \tfrac{21827575}{2426112}c\bigr)- \omega^{2} \rho b u\bigl(\tfrac{7}{36} + \tfrac{35}{54}c + \tfrac{385}{162}c^{2} + \tfrac{6545}{729}c^{3}\bigr) \; + \omega^{2} \rho b^{3} u\bigl(\tfrac{385}{972} + \tfrac{6545}{1458}c\bigr)
\\[-1pt] & - \omega u^{3}\bigl(\tfrac{55}{648} + \tfrac{21505}{50544}c + \tfrac{21827575}{11524032}c^{2}\bigr) \; + \omega b^{2} u^{3}\bigl(\tfrac{623645}{606528}\bigr) \; + \omega^{2} \rho b u^{4}\bigl(\tfrac{385}{2916}\bigr)
\Bigr] +h\Bigl[
\omega^{2} \rho b\bigl(\tfrac{35}{54} + \tfrac{385}{81}c + \tfrac{6545}{243}c^{2}
\\[-1pt] &
+ \tfrac{150535}{729}c^{3}\bigr)- \omega^{2} \rho b^{3}\bigl(\tfrac{9625}{2916} + \tfrac{65450}{729}c\bigr) \; + \omega u^{2}\bigl(\tfrac{51425}{33696} + \tfrac{49267955}{3841344}c + \tfrac{894930575}{7682688}c^{2}\bigr) \; - \omega b^{2} u^{2}\bigl(\tfrac{21827575}{404352}\bigr)
\\[-1pt] & - \omega^{2} \rho b u^{3}\bigl(\tfrac{6545}{729}\bigr)
\Bigr]\Biggr\}+ \frac{(\sqrt3-i)\Gamma\!\bigl(\frac13\bigr)^3}{16\pi^2}\Biggl\{\frac{1}{h}\Bigl[
\omega b\bigl(1 + \tfrac{2}{3}c + \tfrac{14}{9}c^{2} + \tfrac{364}{81}c^{3} + \tfrac{3458}{243}c^{4}\bigr) - \omega b^{3}\bigl(\tfrac{7}{27} + \tfrac{182}{81}c + \tfrac{3458}{243}c^{2}\bigr)\\[-1pt] &  + \omega b^{5}\bigl(\tfrac{728}{1215}\bigr)+ \omega^{2} \rho u\bigl(\tfrac{1}{2} + \tfrac{7}{60}c + \tfrac{1729}{7920}c^{2} + \tfrac{267995}{484704}c^{3} + \tfrac{426380045}{267556608}c^{4}\bigr) - \omega^{2} \rho b^{2} u\bigl(\tfrac{91}{720} + \tfrac{8645}{9504}c + \tfrac{9915815}{1938816}c^{2}\bigr)\\[-1pt] &  + \omega^{2} \rho b^{4} u\bigl(\tfrac{267995}{684288}\bigr) \; - \omega b u^{3}\bigl(\tfrac{14}{81} + \tfrac{364}{243}c\bigr) - \omega^{2} \rho u^{4}\bigl(\tfrac{1729}{142560} + \tfrac{267995}{2908224}c\bigr)
\Bigr] +h\Bigl[
-\omega^{2} \rho\bigl(\tfrac{23}{120} + \tfrac{4277}{7920}c + \tfrac{613795}{323136}c^{2}\\[-1pt] &
+ \tfrac{942002425}{133778304}c^{3} + \tfrac{20892622205}{535113216}c^{4}\bigr) + \omega^{2} \rho b^{2}\bigl(\tfrac{70889}{95040} + \tfrac{17419675}{1938816}c + \tfrac{426380045}{3877632}c^{2}\bigr) \; - \omega^{2} \rho b^{4}\bigl(\tfrac{9915815}{1368576}\bigr) \; \\[-1pt] &
+ \omega b u^{2}\bigl(\tfrac{364}{81} + \tfrac{6916}{81}c\bigr)+ \omega^{2} \rho u^{3}\bigl(\tfrac{613795}{1454112} + \tfrac{9915815}{1454112}c\bigr)
\Bigr]\Biggr\}+\mathcal{O}(6)
\end{align*}

\begin{align*}
&B_3 =\frac{(1-\omega)\sqrt3\,\Gamma\!\bigl(\frac13\bigr)^3}{5\,2^{4/3}\pi^2}\,\frac1h\Biggl[ (-c)^{5/6}\bigl(1-\tfrac9{22}(-c)+\tfrac{945}{1496}(-c)^2
   -\tfrac{8505}{6256}(-c)^3 +\tfrac{4975425}{1451392}(-c)^4
   -\tfrac{1620567}{170752}(-c)^5\bigr)
\\[-1pt]
& + b^2(-c)^{-1/6}\bigl(\tfrac5{24}-\tfrac{15}{16}(-c)+\tfrac{2835}{704}(-c)^2
   -\tfrac{36855}{2176}(-c)^3+\tfrac{1658475}{23552}(-c)^4\bigr)
 - b^4(-c)^{-7/6}\bigl(\tfrac5{1152}+\tfrac{175}{768}(-c)\\[-1pt] &
   -\tfrac{3465}{1024}(-c)^2+\tfrac{61425}{2048}(-c)^3\bigr) + b^6(-c)^{-13/6}\bigl(\tfrac{35}{82944}+\tfrac{35}{6144}(-c)
    +\tfrac{5005}{8192}(-c)^2\bigr)
   - b^8(-c)^{-19/6}\bigl(\tfrac{455}{7962624}\\[-1pt] &+\tfrac{385}{589824}(-c)\bigr) + \tfrac{1729}{191102976}b^{10}(-c)^{-25/6}
    + u^3(-c)^{-7/6}\bigl(\tfrac5{648}+\tfrac5{144}(-c)-\tfrac{35}{192}(-c)^2\bigr) - b^2u^3(-c)^{-13/6}\bigl(\tfrac{35}{15552}\\[-1pt] &+\tfrac{25}{3456}(-c)\bigr)
    + \tfrac{455}{746496}b^4u^3(-c)^{-19/6}\Biggr]
-\frac{(1-\omega^2)2^{4/3}\pi}{3\,\Gamma\!\bigl(\frac13\bigr)^3}\Biggl\{\frac1h\Bigl[ u(-c)^{1/6}\bigl(1-\tfrac9{14}(-c)+\tfrac{135}{104}(-c)^2
   -\tfrac{13365}{3952}(-c)^3 \\[-1pt] &+\tfrac{24057}{2432}(-c)^4\bigr) +b^2u(-c)^{-5/6}\bigl(\tfrac1{24}-\tfrac{15}{16}(-c)
   +\tfrac{405}{64}(-c)^2-\tfrac{4455}{128}(-c)^3\bigr)
 -b^4u(-c)^{-11/6}\bigl(\tfrac5{1152}+\tfrac{35}{768}(-c)
   \\[-1pt] &-\tfrac{3465}{1024}(-c)^2\bigr)
   +b^6u(-c)^{-17/6}\bigl(\tfrac{55}{82944}+\tfrac{35}{6144}(-c)\bigr) -\tfrac{935}{7962624}b^8u(-c)^{-23/6}
   +u^4(-c)^{-11/6}\bigl(\tfrac5{1296}+\tfrac1{288}(-c)\bigr)
   \\[-1pt] &-\tfrac{55}{31104}b^2u^4(-c)^{-17/6}\Bigr] +h\Bigl[-\tfrac1{12}(-c)^{-5/6}-\tfrac98(-c)^{1/6}
    +\tfrac5{288}b^2(-c)^{-11/6}\Bigr]\Biggr\}+\mathcal{O}(6)
\end{align*}

\begin{align*}
&\mathcal A_1 =2\pi i\,\frac{6\pi\bigl(i+\frac{1}{\sqrt3}\bigr)}{7\,2^{1/3}\Gamma\!\bigl(\frac13\bigr)^3}\Biggl\{\frac{1}{h}\Bigl[
\omega\bigl(1 + \tfrac{7}{6}c + \tfrac{55}{72}c^{2} + \tfrac{21505}{16848}c^{3} + \tfrac{21827575}{7682688}c^{4} + \tfrac{1682469481}{230480640}c^{5}\bigr) - \omega b^{2}\bigl(\tfrac{35}{72} + \tfrac{935}{432}c\\[-1pt] & + \tfrac{623645}{67392}c^{2} + \tfrac{894930575}{23048064}c^{3}\bigr) \; + \omega b^{4}\bigl(\tfrac{21505}{31104} + \tfrac{21827575}{2426112}c\bigr) + \omega^{2} \rho b u\bigl(\tfrac{7}{36} + \tfrac{35}{54}c + \tfrac{385}{162}c^{2} + \tfrac{6545}{729}c^{3}\bigr) \\[-1pt] & - \omega^{2} \rho b^{3} u\bigl(\tfrac{385}{972} + \tfrac{6545}{1458}c\bigr) - \omega u^{3}\bigl(\tfrac{55}{648} + \tfrac{21505}{50544}c + \tfrac{21827575}{11524032}c^{2}\bigr) \; + \omega b^{2} u^{3}\bigl(\tfrac{623645}{606528}\bigr) \; - \omega^{2} \rho b u^{4}\bigl(\tfrac{385}{2916}\bigr)
\Bigr]
\\[2pt] & +h\Bigl[
-\omega^{2} \rho b\bigl(\tfrac{35}{54} + \tfrac{385}{81}c + \tfrac{6545}{243}c^{2} + \tfrac{150535}{729}c^{3}\bigr) + \omega^{2} \rho b^{3}\bigl(\tfrac{9625}{2916} + \tfrac{65450}{729}c\bigr) \; + \omega u^{2}\bigl(\tfrac{51425}{33696} + \tfrac{49267955}{3841344}c \\[-1pt] &+ \tfrac{894930575}{7682688}c^{2}\bigr) \; - \omega b^{2} u^{2}\bigl(\tfrac{21827575}{404352}\bigr) + \omega^{2} \rho b u^{3}\bigl(\tfrac{6545}{729}\bigr)
\Bigr]\Biggr\}+ 2\pi i\,\frac{(\sqrt3-i)\Gamma\!\bigl(\frac13\bigr)^3}{16\pi^2}\Biggl\{\frac{1}{h}\Bigl[
-\omega b\bigl(1 + \tfrac{2}{3}c + \tfrac{14}{9}c^{2}\\[-1pt] & + \tfrac{364}{81}c^{3} + \tfrac{3458}{243}c^{4}\bigr)
 + \omega b^{3}\bigl(\tfrac{7}{27} + \tfrac{182}{81}c + \tfrac{3458}{243}c^{2}\bigr)  - \omega b^{5}\bigl(\tfrac{728}{1215}\bigr) + \omega^{2} \rho u\bigl(\tfrac{1}{2} + \tfrac{7}{60}c + \tfrac{1729}{7920}c^{2} + \tfrac{267995}{484704}c^{3}\\[-1pt] & + \tfrac{426380045}{267556608}c^{4}\bigr) - \omega^{2} \rho b^{2} u\bigl(\tfrac{91}{720} + \tfrac{8645}{9504}c + \tfrac{9915815}{1938816}c^{2}\bigr)  + \omega^{2} \rho b^{4} u\bigl(\tfrac{267995}{684288}\bigr) \; + \omega b u^{3}\bigl(\tfrac{14}{81} + \tfrac{364}{243}c\bigr)\\[-1pt] & - \omega^{2} \rho u^{4}\bigl(\tfrac{1729}{142560} + \tfrac{267995}{2908224}c\bigr)
\Bigr] +h\Bigl[
-\omega^{2} \rho\bigl(\tfrac{23}{120} + \tfrac{4277}{7920}c + \tfrac{613795}{323136}c^{2} + \tfrac{942002425}{133778304}c^{3} + \tfrac{20892622205}{535113216}c^{4}\bigr)
\\[-1pt] & + \omega^{2} \rho b^{2}\bigl(\tfrac{70889}{95040} + \tfrac{17419675}{1938816}c + \tfrac{426380045}{3877632}c^{2}\bigr) \; - \omega^{2} \rho b^{4}\bigl(\tfrac{9915815}{1368576}\bigr) \; - \omega b u^{2}\bigl(\tfrac{364}{81} + \tfrac{6916}{81}c\bigr)
\\[-1pt] & + \omega^{2} \rho u^{3}\bigl(\tfrac{613795}{1454112} + \tfrac{9915815}{1454112}c\bigr)
\Bigr]\Biggr\}+\mathcal{O}(6)
\end{align*}

\begin{align*}
&\mathcal A_2 =-2\pi i\,\frac{6\pi\bigl(i+\frac{1}{\sqrt3}\bigr)}{7\,2^{1/3}\Gamma\!\bigl(\frac13\bigr)^3}\Biggl\{\frac{1}{h}\Bigl[
1 + \tfrac{7}{6}c + \tfrac{55}{72}c^{2} + \tfrac{21505}{16848}c^{3} + \tfrac{21827575}{7682688}c^{4} + \tfrac{1682469481}{230480640}c^{5} - b^{2}\bigl(\tfrac{35}{72} + \tfrac{935}{432}c + \tfrac{623645}{67392}c^{2} \\[-1pt] &+ \tfrac{894930575}{23048064}c^{3}\bigr) \; + b^{4}\bigl(\tfrac{21505}{31104} + \tfrac{21827575}{2426112}c\bigr)+ \rho b u\bigl(\tfrac{7}{36} + \tfrac{35}{54}c + \tfrac{385}{162}c^{2} + \tfrac{6545}{729}c^{3}\bigr) \; - \rho b^{3} u\bigl(\tfrac{385}{972} + \tfrac{6545}{1458}c\bigr)
\\[-1pt] & - u^{3}\bigl(\tfrac{55}{648} + \tfrac{21505}{50544}c + \tfrac{21827575}{11524032}c^{2}\bigr) \; + b^{2} u^{3}\bigl(\tfrac{623645}{606528}\bigr) \; - \rho b u^{4}\bigl(\tfrac{385}{2916}\bigr)
\Bigr]+h\Bigl[
-\rho b\bigl(\tfrac{35}{54} + \tfrac{385}{81}c + \tfrac{6545}{243}c^{2} + \tfrac{150535}{729}c^{3}\bigr)
\\[-1pt] & + \rho b^{3}\bigl(\tfrac{9625}{2916} + \tfrac{65450}{729}c\bigr) \; + u^{2}\bigl(\tfrac{51425}{33696} + \tfrac{49267955}{3841344}c + \tfrac{894930575}{7682688}c^{2}\bigr) \; - b^{2} u^{2}\bigl(\tfrac{21827575}{404352}\bigr) + \rho b u^{3}\bigl(\tfrac{6545}{729}\bigr)
\Bigr]\Biggr\}\\[-1pt] &- 2\pi i\,\frac{(\sqrt3-i)\Gamma\!\bigl(\frac13\bigr)^3}{16\pi^2}\Biggl\{\frac{1}{h}\Bigl[
-b\bigl(1 + \tfrac{2}{3}c + \tfrac{14}{9}c^{2} + \tfrac{364}{81}c^{3} + \tfrac{3458}{243}c^{4}\bigr) + b^{3}\bigl(\tfrac{7}{27} + \tfrac{182}{81}c + \tfrac{3458}{243}c^{2}\bigr) \; - b^{5}\bigl(\tfrac{728}{1215}\bigr)
\\[-1pt] & + \rho u\bigl(\tfrac{1}{2} + \tfrac{7}{60}c + \tfrac{1729}{7920}c^{2} + \tfrac{267995}{484704}c^{3} + \tfrac{426380045}{267556608}c^{4}\bigr) \; - \rho b^{2} u\bigl(\tfrac{91}{720} + \tfrac{8645}{9504}c + \tfrac{9915815}{1938816}c^{2}\bigr) + \rho b^{4} u\bigl(\tfrac{267995}{684288}\bigr)\\[-1pt] & + b u^{3}\bigl(\tfrac{14}{81} + \tfrac{364}{243}c\bigr) \; - \rho u^{4}\bigl(\tfrac{1729}{142560} + \tfrac{267995}{2908224}c\bigr)
\Bigr]+h\Bigl[
-\rho\bigl(\tfrac{23}{120} + \tfrac{4277}{7920}c + \tfrac{613795}{323136}c^{2} + \tfrac{942002425}{133778304}c^{3} + \tfrac{20892622205}{535113216}c^{4}\bigr)
\\[-1pt] & + \rho b^{2}\bigl(\tfrac{70889}{95040} + \tfrac{17419675}{1938816}c + \tfrac{426380045}{3877632}c^{2}\bigr) \; - \rho b^{4}\bigl(\tfrac{9915815}{1368576}\bigr) \; - b u^{2}\bigl(\tfrac{364}{81} + \tfrac{6916}{81}c\bigr) + \rho u^{3}\bigl(\tfrac{613795}{1454112} + \tfrac{9915815}{1454112}c\bigr)
\Bigr]\Biggr\}\\[-1pt] &+\mathcal{O}(6)
\end{align*}

\begin{align*}
&\mathcal A_3 =-2\pi i\omega\,\frac{(1-\omega)\sqrt3\,\Gamma\!\bigl(\frac13\bigr)^3}{5\,2^{4/3}\pi^2}\,\frac1h\Biggl[
 (-c)^{5/6}\bigl(1-\tfrac9{22}(-c)+\tfrac{945}{1496}(-c)^2
   -\tfrac{8505}{6256}(-c)^3 +\tfrac{4975425}{1451392}(-c)^4
   -\tfrac{1620567}{170752}(-c)^5\bigr)
\\[-1pt]
& + b^2(-c)^{-1/6}\bigl(\tfrac5{24}-\tfrac{15}{16}(-c)+\tfrac{2835}{704}(-c)^2
   -\tfrac{36855}{2176}(-c)^3+\tfrac{1658475}{23552}(-c)^4\bigr)
 - b^4(-c)^{-7/6}\bigl(\tfrac5{1152}+\tfrac{175}{768}(-c)\\[-1pt] &
   -\tfrac{3465}{1024}(-c)^2+\tfrac{61425}{2048}(-c)^3\bigr)
 + b^6(-c)^{-13/6}\bigl(\tfrac{35}{82944}+\tfrac{35}{6144}(-c)
    +\tfrac{5005}{8192}(-c)^2\bigr)
   - b^8(-c)^{-19/6}\bigl(\tfrac{455}{7962624}\\[-1pt] &+\tfrac{385}{589824}(-c)\bigr) + \tfrac{1729}{191102976}b^{10}(-c)^{-25/6}
    + u^3(-c)^{-7/6}\bigl(\tfrac5{648}+\tfrac5{144}(-c)-\tfrac{35}{192}(-c)^2\bigr) - b^2u^3(-c)^{-13/6}\bigl(\tfrac{35}{15552}\\[-1pt] &+\tfrac{25}{3456}(-c)\bigr)
    + \tfrac{455}{746496}b^4u^3(-c)^{-19/6}\Biggr]+2\pi i\omega^2\,\frac{(1-\omega^2)2^{4/3}\pi}{3\,\Gamma\!\bigl(\frac13\bigr)^3}\Biggl\{\frac1h\Bigl[
 u(-c)^{1/6}\bigl(1-\tfrac9{14}(-c)+\tfrac{135}{104}(-c)^2
   \\[-1pt] &-\tfrac{13365}{3952}(-c)^3+\tfrac{24057}{2432}(-c)^4\bigr) +b^2u(-c)^{-5/6}\bigl(\tfrac1{24}-\tfrac{15}{16}(-c)
   +\tfrac{405}{64}(-c)^2-\tfrac{4455}{128}(-c)^3\bigr) -b^4u(-c)^{-11/6}\bigl(\tfrac5{1152}\\[-1pt] &+\tfrac{35}{768}(-c)
   -\tfrac{3465}{1024}(-c)^2\bigr)
   +b^6u(-c)^{-17/6}\bigl(\tfrac{55}{82944}+\tfrac{35}{6144}(-c)\bigr) -\tfrac{935}{7962624}b^8u(-c)^{-23/6}
   +u^4(-c)^{-11/6}\bigl(\tfrac5{1296}\\[-1pt] &+\tfrac1{288}(-c)\bigr)
   -\tfrac{55}{31104}b^2u^4(-c)^{-17/6}\Bigr] +h\Bigl[-\tfrac1{12}(-c)^{-5/6}-\tfrac98(-c)^{1/6}
    +\tfrac5{288}b^2(-c)^{-11/6}\Bigr]\Biggr\}+\mathcal{O}(6)
\end{align*}
\endgroup

\bibliography{refs}

\providecommand{\href}[2]{#2}\begingroup\raggedright\begin{thebibliography}{10}

\bibitem{Bender:1969si}
C.~M. Bender and T.~T. Wu, {\it {Anharmonic oscillator}},  {\em Phys. Rev.}
  {\bf 184} (1969) 1231--1260.

\bibitem{Ecalle:1981}
J.~\'Ecalle, {\em {Les Fonctions R\'esurgentes}}, vol.~I-III.
\newblock Publications Math\'ematiques d'Orsay, Orsay, France, 1981.

\bibitem{DDP}
E.~Delabaere, H.~Dillinger, and F.~Pham, {\it {Exact semiclassical expansions
  for one-dimensional quantum oscillators}},  {\em J. Math. Phys. (N.Y.)} {\bf
  38} (12, 1997) 6126--6184.

\bibitem{Voros1983}
A.~Voros, {\it {The return of the quartic oscillator. The complex WKB method}},
   {\em Annales de l'I.H.P. Physique théorique} {\bf 39} (1983), no.~3
  211--338.

\bibitem{Aniceto:2013fka}
I.~Aniceto and R.~Schiappa, {\it {Nonperturbative Ambiguities and the Reality
  of Resurgent Transseries}},  {\em Commun. Math. Phys.} {\bf 335} (2015),
  no.~1 183--245, [\href{http://arxiv.org/abs/1308.1115}{{\tt
  arXiv:1308.1115}}].

\bibitem{Aniceto:2018bis}
I.~Aniceto, G.~Ba\c{s}ar, and R.~Schiappa, {\it {A primer on resurgent
  transseries and their asymptotics}},  {\em Phys. Rep.} {\bf 809} (2019)
  1--135, [\href{http://arxiv.org/abs/1802.10441}{{\tt arXiv:1802.10441}}].

\bibitem{Zinn-Justin:2004vcw}
J.~Zinn-Justin and U.~D. Jentschura, {\it {Multi-instantons and exact results
  I: Conjectures, WKB expansions, and instanton interactions}},  {\em Ann.
  Phys. (N.Y.)} {\bf 313} (2004) 197--267,
  [\href{http://arxiv.org/abs/quant-ph/0501136}{{\tt quant-ph/0501136}}].

\bibitem{Zinn-Justin:2004qzw}
J.~Zinn-Justin and U.~D. Jentschura, {\it {Multi-instantons and exact results
  II: Specific cases, higher-order effects, and numerical calculations}},  {\em
  Ann. Phys. (N.Y.)} {\bf 313} (2004) 269--325,
  [\href{http://arxiv.org/abs/quant-ph/0501137}{{\tt quant-ph/0501137}}].

\bibitem{Jentschura:2010zza}
U.~D. Jentschura, A.~Surzhykov, and J.~Zinn-Justin, {\it {Multi-instantons and
  exact results III: Unification of even and odd anharmonic oscillators}},
  {\em Ann. Phys. (N.Y.)} {\bf 325} (2010) 1135--1172,
  [\href{http://arxiv.org/abs/1001.3910}{{\tt arXiv:1001.3910}}].

\bibitem{Codesido:2017jwp}
S.~Codesido, M.~Marino, and R.~Schiappa, {\it {Non-Perturbative Quantum
  Mechanics from Non-Perturbative Strings}},  {\em Annales Henri Poincare} {\bf
  20} (2019), no.~2 543--603, [\href{http://arxiv.org/abs/1712.02603}{{\tt
  arXiv:1712.02603}}].

\bibitem{vanSpaendonck:2023znn}
A.~van Spaendonck and M.~Vonk, {\it {Exact instanton transseries for quantum
  mechanics}},  {\em SciPost Phys.} {\bf 16} (2024), no.~4 103,
  [\href{http://arxiv.org/abs/2309.05700}{{\tt arXiv:2309.05700}}].

\bibitem{Gu:2022fss}
J.~Gu and M.~Marino, {\it {On the resurgent structure of quantum periods}},
  {\em SciPost Phys.} {\bf 15} (2023), no.~1 035,
  [\href{http://arxiv.org/abs/2211.03871}{{\tt arXiv:2211.03871}}].

\bibitem{Alvarez:2000}
G.~Álvarez and C.~Casares, {\it Exponentially small corrections in the
  asymptotic expansion of the eigenvalues of the cubic anharmonic oscillator},
  {\em J. Phys. A} {\bf 33} (2000), no.~29 5171.

\bibitem{Alvarez2004}
G.~\'{A}lvarez, {\it {Langer–Cherry derivation of the multi-instanton
  expansion for the symmetric double well}},  {\em J. Math. Phys. (N.Y.)} {\bf
  45} (07, 2004) 3095--3108.

\bibitem{Dunne:2013ada}
G.~V. Dunne and M.~\"Unsal, {\it {Generating nonperturbative physics from
  perturbation theory}},  {\em Phys. Rev. D} {\bf 89} (2014), no.~4 041701,
  [\href{http://arxiv.org/abs/1306.4405}{{\tt arXiv:1306.4405}}].

\bibitem{Dunne:2014bca}
G.~V. Dunne and M.~{\"U}nsal, {\it {Uniform WKB, multi-instantons, and
  resurgent trans-series}},  {\em Phys. Rev. D} {\bf 89} (2014), no.~10 105009,
  [\href{http://arxiv.org/abs/1401.5202}{{\tt arXiv:1401.5202}}].

\bibitem{Basar:2017hpr}
G.~Ba\c{s}ar, G.~V. Dunne, and M.~{\"U}nsal, {\it {Quantum geometry of
  resurgent perturbative/nonperturbative relations}},  {\em J. High Energy
  Phys.} {\bf 2017} (2017), no.~5 87,
  [\href{http://arxiv.org/abs/1701.06572}{{\tt arXiv:1701.06572}}].

\bibitem{Gorsky:2014lia}
A.~Gorsky and A.~Milekhin, {\it {RG-Whitham dynamics and complex Hamiltonian
  systems}},  {\em Nucl. Phys. B} {\bf 895} (2015) 33--63,
  [\href{http://arxiv.org/abs/1408.0425}{{\tt arXiv:1408.0425}}].

\bibitem{Kozcaz:2026isf}
C.~Koz{\c{c}}az, {\it {Quantum Residues and the Geometry of
  Perturbative/Non-perturbative Relations}},
  \href{http://arxiv.org/abs/2610.19258}{{\tt arXiv:2610.19258}}.

\bibitem{Cavusoglu:2023bai}
A.~\c{C}avu\c{s}o\u{g}lu, C.~Koz\c{c}az, and K.~Tezgin, {\it {Unified genus-1
  potential and a parametric perturbative/nonperturbative relation}},  {\em
  Phys. Rev. D} {\bf 110} (2024), no.~12 125001,
  [\href{http://arxiv.org/abs/2311.17850}{{\tt arXiv:2311.17850}}].

\bibitem{Cavusoglu:2024usn}
A.~{\c{C}}avu{\c{s}}o{\u{g}}lu, C.~Koz{\c{c}}az, and K.~Tezgin, {\it
  {Resurgence of deformed genus-1 curves: A novel perturbative/nonperturbative
  relation}},  {\em Phys. Rev. D} {\bf 111} (2025), no.~10 105007,
  [\href{http://arxiv.org/abs/2408.02628}{{\tt arXiv:2408.02628}}].

\bibitem{Gahramanov:2015yxk}
I.~Gahramanov and K.~Tezgin, {\it {Remark on the Dunne-\"Unsal relation in
  exact semiclassics}},  {\em Phys. Rev. D} {\bf 93} (2016), no.~6 065037,
  [\href{http://arxiv.org/abs/1512.08466}{{\tt arXiv:1512.08466}}].

\bibitem{Gahramanov:2016xjj}
I.~Gahramanov and K.~Tezgin, {\it {A resurgence analysis for cubic and quartic
  anharmonic potentials}},  {\em Int. J. Mod. Phys. A} {\bf 32} (2017), no.~05
  1750033, [\href{http://arxiv.org/abs/1608.08119}{{\tt arXiv:1608.08119}}].

\bibitem{Fischbach:2018yiu}
F.~Fischbach, A.~Klemm, and C.~Nega, {\it {WKB method and quantum periods
  beyond genus one}},  {\em J. Phys. A} {\bf 52} (2019), no.~7 075402,
  [\href{http://arxiv.org/abs/1803.11222}{{\tt arXiv:1803.11222}}].

\bibitem{Ture:2026ktb}
M.~T{\"u}re and M.~{\"U}nsal, {\it {Quantum geometry at arbitrary genus - I:
  Anharmonic potentials}},  \href{http://arxiv.org/abs/2610.10940}{{\tt
  arXiv:2610.10940}}.

\bibitem{Gah}
I.~Gahramanov, {\it {Remark on perturbative/non-perturbative relations at
  higher genus}}, .

\bibitem{AT}
A.~{\c{C}}avu{\c{s}}o{\u{g}}lu and K.~Tezgin, {\it {Constructive Resurgence in
  Higher Genus: Quantum Geometry and Moduli Spaces of WKB Curves}}, .

\bibitem{Rauch}
H.~E. Rauch, {\it {Weierstrass points, branch points, and moduli of Riemann
  surfaces}},  {\em Commun. Pure Appl. Math.} {\bf 12} (1959) 543.

\bibitem{Eynard:2007kz}
B.~Eynard and N.~Orantin, {\it {Invariants of algebraic curves and topological
  expansion}},  {\em Commun. Num. Theor. Phys.} {\bf 1} (2007) 347--452,
  [\href{http://arxiv.org/abs/math-ph/0702045}{{\tt math-ph/0702045}}].

\bibitem{Chekhov:2010zg}
L.~O. Chekhov, B.~Eynard, and O.~Marchal, {\it {Topological expansion of
  $\beta$-ensemble model and quantum algebraic geometry in the sectorwise
  approach}},  {\em Theor. Math. Phys.} {\bf 166} (2011) 141--185,
  [\href{http://arxiv.org/abs/1009.6007}{{\tt arXiv:1009.6007}}].

\bibitem{Kozcaz:2016wvy}
C.~Koz\c{c}az, T.~Sulejmanpasic, Y.~Tanizaki, and M.~\"Unsal, {\it {Cheshire
  Cat resurgence, self-resurgence and quasi-exact solvable systems}},  {\em
  Commun. Math. Phys.} {\bf 364} (2018), no.~3 835--878,
  [\href{http://arxiv.org/abs/1609.06198}{{\tt arXiv:1609.06198}}].

\bibitem{Kamata:2021jrs}
S.~Kamata, T.~Misumi, N.~Sueishi, and M.~{\"U}nsal, {\it {Exact WKB analysis
  for SUSY and quantum deformed potentials: Quantum mechanics with Grassmann
  fields and Wess-Zumino terms}},  {\em Phys. Rev. D} {\bf 107} (2023), no.~4
  045019, [\href{http://arxiv.org/abs/2111.05922}{{\tt arXiv:2111.05922}}].

\bibitem{Koike}
K.~T., {\it {On the exact WKB analysis of second order linear ordinary
  differential equations with simple poles}},  {\em Publ. RIMS Kyoto Univ.}
  {\bf 36} (2000) 297.

\bibitem{Ito:2021sjo}
K.~Ito, T.~Kondo, and H.~Shu, {\it {Wall-crossing of TBA equations and WKB
  periods for the third order ODE}},  {\em Nucl. Phys. B} {\bf 979} (2022)
  115788, [\href{http://arxiv.org/abs/2111.11047}{{\tt arXiv:2111.11047}}].

\end{thebibliography}\endgroup

\end{document}